\documentclass[twocolumn,trackchanges,twocolappendix]{aastex7}
\received{Mar 28, 2025}
\accepted{Aug 17, 2026}
\submitjournal{ApJ}
\usepackage{mysymbol}
\usepackage{chemformula}
\let\ce\ch
\usepackage{comment}
\usepackage{threeparttable}

\usepackage[T1]{fontenc}

\newcommand{\htwoco}{$\mathrm{H_2 CO}$}
\newcommand{\hcop}{$\mathrm{HCO^{+}}$}

\begin{document}

\title{Nonthermal Velocity Dispersion in the Outer Disk of HL Tau}

\author[0000-0003-4361-5577]{Jinshi Sai}
\affiliation{Department of Physics and Astronomy, Graduate School of Science and Engineering, Kagoshima University, 1-21-35 Korimoto, Kagoshima, Kagoshima 890-0065, Japan}
\email{jn.insa.sai@gmail.com}
\email[show]{jinshi.sai@sci.kagoshima-u.ac.jp}

\author[0000-0003-0845-128X]{Shigehisa Takakuwa}
\affiliation{Department of Physics and Astronomy, Graduate School of Science and Engineering, Kagoshima University, 1-21-35 Korimoto, Kagoshima, Kagoshima 890-0065, Japan}
\affiliation{Academia Sinica Institute of Astronomy \& Astrophysics,
11F of Astronomy-Mathematics Building, AS/NTU, No.1, Sec. 4, Roosevelt
Rd, Taipei 106319, Taiwan}
\email{takakuwa@sci.kagoshima-u.ac.jp}

\author[0000-0003-1412-893X]{Hsi-Wei Yen}
\affiliation{Academia Sinica Institute of Astronomy \& Astrophysics,
11F of Astronomy-Mathematics Building, AS/NTU, No.1, Sec. 4, Roosevelt
Rd, Taipei 106319, Taiwan}
\email{hwyen@asiaa.sinica.edu.tw}

\author[0000-0001-6738-676X]{Yusuke Tsukamoto}
\affiliation{Department of Physics and Astronomy, Graduate School of Science and Engineering, Kagoshima University, 1-21-35 Korimoto,
Kagoshima, Kagoshima 890-0065, Japan}
\affiliation{Department of Physics, Faculty of Science and Engineering, Konan University, Okamoto 8-9-1, Higashinada-ku, Kobe 658-8501, Japan}
\email{tsukamoto.yusuke@sci.kagoshima-u.ac.jp}

\author[0000-0002-9660-8947]{Yuya Fukuhara}
\affiliation{Academia Sinica Institute of Astronomy \& Astrophysics,
11F of Astronomy-Mathematics Building, AS/NTU, No.1, Sec. 4, Roosevelt
Rd, Taipei 106319, Taiwan}
\email{yfukuhara@asiaa.sinica.edu.tw}

\begin{abstract}

Turbulence in protoplanetary disks plays a crucial role in the evolution of disk structures and the planet formation process. However, the strength of the turbulence remains unclear in young, embedded disks surrounded by infalling envelopes. In this paper, we present the first direct measurement of the nonthermal velocity dispersion within the embedded disk around HL Tau, which possesses a dusty disk with multiple rings and gaps but is still associated with infalling gas flows from an envelope. Using ALMA archival data of the \ch{H_2CO} emission, we measured the local line width through parametric model fitting that accounts for the contribution of Keplerian shear motion. After subtracting the thermal component, the nonthermal velocity dispersion is estimated to be $\rtsim0.12\mbox{--}0.15~\kmps$ on average over radii of $80\mbox{--}180~\au$, and it slightly increases with radius. The inferred nonthermal motions correspond to a turbulent mach number of $\mathcal{M}\tsim0.4$ or a viscous $\alpha$ value of $\alpha \tsim0.16$, assuming that it is entirely caused by turbulence and $\alpha\tsim \mathcal{M}^2$. In addition, fitting with models with elevated molecular layers suggests that the \ch{H_2CO} emission primarily traces near the disk midplane ($z/R \tsim0.05$). Turbulence driven by gravitational instability or infall from the envelope most naturally explains the large nonthermal motions, considering the large disk mass and associated infalling streamers. The strong turbulence measured in the outer disk, in contrast to the vertically settled inner dusty disk, suggests a pronounced radial variation in the turbulence strength and/or an anisotropic nature of the turbulence within the disk.
\end{abstract}


%\keywords{\uat{Protoplanetary disk}{xxx} --- \uat{Disk turbulence}{343} --- \uat{HL Tau}{739}}

\section{Introduction} \label{sec:intro}

% General introduction
Protoplanetary disks are expected to be turbulent due to the magneto-rotational instability or hydrodynamic instabilities \citep{Lesur2023a}. The strength of turbulence is a crucial parameter in planet formation models. The evolution of solid particles within disks depends on the turbulence strength \citep{Birnstiel2016a}. Planetesimal formation via the streaming instability requires the concentration of dust grains that have settled toward the disk midplane, whereas strong turbulent mixing inhibits this process \citep{Drazkowska2018a}. The turbulence in disks does not only affect dynamics of solid particles, but also influence chemical processes within disks \citep{Furuya2014a}. Therefore, constraining strength of turbulence and its driving mechanisms in disks is crucial for understanding of disk evolution and planet formation processes.

% Observational progress so far
Turbulence in protoplanetary disks is described by the dimensionless parameter $\alpha$ in the context of the viscous $\alpha$-disk model \citep{Shakura1973a}. In this framework, the kinetic viscosity $\nu$, as a product of turbulence, is expressed with $\alpha$ as $\nu = \alpha \cs H$, where $\cs$ is the isothermal sound speed, and $H$ is the gas pressure scale height. Although deriving $\alpha$ directly from observations is difficult, there are a few ways to infer $\alpha$ through a conversion from observed quantities \citep{Rosotti2023a}.

Several studies have observationally investigated the strength of turbulence on scales of $\rtsim30$--$500~\au$ in disks in Class \II~sources (or T Tauri stars; $t_\mathrm{age}\!\!\sim\!1$--$5~\mathrm{Myr}$). The most direct way is to measure the turbulent line width with molecular line observations. Direct measurements of nonthermal broadening of line emission result in a typical upper limit of $\rtsim5\mbox{--}10\%$ of the isothermal sound speed in four disks \citep[e.g.,][]{Hughes2011a, Flaherty2017a, Flaherty2018a, Flaherty2020a, Teague2016a, Teague2018c}, and detections of strong turbulence ($\vturb \sim 0.2\mbox{--}0.7~\cs$) in two disks of DM Tau and IM Lup \citep{Guilloteau2012a, Flaherty2020a, Paneque-Carreno2024a, Flaherty2024a, Hardiman2026a}. These quantities can be equated to the dimensionless parameter $\alpha$ assuming $\alpha \sim \mathcal{M}^2 = (\delta v_\mathrm{turb}/\cs/\sqrt{2})^2$ \citep{Pinte2022b}, although it should be noted that this conversion is not necessarily always valid \citep{Lesur2023a}. These estimated turbulent velocities correspond to $\alpha \lesssim 10^{-2}$--$10^{-3}$ in the weak turbulence case and $\alpha\sim 0.1$ in the strong turbulence case.

The vertical thickness of the dusty disks has also been used as an indirect way to probe the strength of turbulence via dust diffusion \citep{Pinte2016a}. With this method, dust scale heights of $\Hd/R < 0.05$ much smaller than typical gas pressure scale heights were suggested in the Class \II~disk of Oph 163131, which correspond to $\alpha \tsim 10^{-4}$ or less \citep{Villenave2022a}. With similar methods, the dust scale heights were estimated in the disk of HD 163296, and moderate to strong turbulence of $\alpha \lesssim 10^{-3}$ and $\alpha \gtrsim 10^{-2}$ was inferred with a radial variation \citep[e.g.,][]{Ohashi2019a, Doi2021a, Doi2023a}. More recent studies expand these analyses to a sample of Class \II~disks, reporting that turbulence is typically $\alpha \lesssim 10^{-4}\mbox{--}10^{-3}$ in outer disks \citep{Villenave2025a, Jiang2025a}.

% about protostars
While these previous studies offer strong constraints on the strength of turbulence in Class \II~disks, the turbulence strength in even younger disks around Class 0 or I protostars ($t_\mathrm{age}\!\!\lesssim\!0.1$--$0.5~\mathrm{Myr}$) remains unclear. Recent discoveries of substructures in these young disks hint at planet formation at this early evolutionary stage \citep[e.g.,][]{Sheehan2020b, Ohashi2023a}, emphasizing the importance of the disks around protostars as the site of planet formation. Therefore, it is crucial to measure turbulence in such protostellar disks.

HL Tau is a nearby protostar located in the Taurus molecular cloud \citep[$d\tsim147~\pc$;][]{Galli2018a}. It has been considered as a Class I/\II~source, i.e., being at a late accretion phase, and is still embedded in an infalling envelope \citep[][]{Mullin2024a}. This source also exhibits a strong molecular jet and outflow \citep{Cabrit1996a, Takami2007a}, and is associated with anisotropic inflows connected to the disk on a scale of $\rtsim300~\au$, as seen in the \ch{HCO^+} (3--2) emission \citep{Yen2019a}. HL Tau possesses a dusty disk showing clear multiple gap and ring structures \citep{ALMAPartnership2015b}. A previous study conducted by \cite{Pinte2016a} found that the dust scale height is as small as $\Hd/R < 0.01$ and estimated the strength of turbulence within the dusty rings to be $\alpha\tsim3\times 10^{-4}$. More recently, \cite{Yang2025a} used the polarized dust emission and derived similar small dust scale heights at radii of $\rtsim40\mbox{--}120~\au$.

In this work, we evaluate the strength of turbulence in the outer disk ($r \gtrsim 80~\au$) of HL Tau using ALMA archival data of the \htwoco~line through measurements of the nonthermal line broadening. The two methods using the dust scale height and the nonthermal motion are complementary because the dust emission is generally sensitive to the inner disk while the molecular line can trace the outer portion of the disk. Measurements of the nonthermal velocity dispersion in embedded disks are challenging due to small gaseous disk size and central stellar mass, and contamination of infalling envelopes and molecular outflows. However, the relatively large disk with a moderate inclination angle of $i\tsim47^\circ$ \citep{ALMAPartnership2015b} and its large protostellar mass make HL Tau a suitable target for the first precise measurement of nonthermal velocity dispersions in embedded disks. Previous observations revealed spatial and velocity structures of several molecular lines in HL Tau \citep{Garufi2021a, Garufi2022a}. In these observations, the \htwoco~line is found to isolate the disk from the infalling envelope, while other molecular lines, such as CO, \hcop~and CS, trace the envelope or outflow. By utilizing the \ch{H_2CO} emission from the HL Tau disk, we demonstrates the capability of measurements of the nonthermal velocity dispersion toward protostellar disks with a reasonable selection of the molecular tracer.

This paper is structured as follows. Section \ref{sec:obs} presents the data reduction process and observational data. Section \ref{sec:modeling} describes the parametric model fitting approach. Section \ref{sec:fitres} presents the fitting results and examines the \ch{H_2CO} emission height. Section \ref{sec:discussion} discusses the interpretation of the results, including potential sources of uncertainties and possible physical mechanisms that can explain the observed nonthermal velocity dispersions. Section \ref{sec:conclusion} summarizes the conclusions of this paper. Additionally, Appendix discusses the accuracy and robustness of the model fitting approach.

\section{Observational Data} \label{sec:obs}

\subsection{Data Reduction}

We analyzed ALMA archival data of the \ce{H2CO} ($3_{1,2}\mbox{--}2_{1,1}$) line (225.697775 GHz; CDMS) of HL Tau (2018.1.01037.S; PI: L.~Podio) \citep{Garufi2021a, Garufi2022a}. The details of the observations were presented in \cite{Garufi2021a}. Observational setup is briefly summarized below. The data were taken in Frequency Division Mode (FDM) during 2018 October 28 to November 3. The 1.2 mm continuum emission was observed within a spectral window with a bandwidth of $1.875~\ghz$ and spectral resolution of $0.976~\mhz$. The \ce{H2CO} ($3_{1,2}\mbox{--}2_{1,1}$) line was observed in a spectral window with a bandwidth of $58.6~\mhz$ and a channel spacing of $122~\khz$, down to which the original channel spacing was reduced with a channel averaging factor of two. This correlator setup resulted in a velocity resolution of $\rtsim0.187~\kmps$ at the frequency of the \ce{H_2CO} line. The data were calibrated with the standard ALMA pipeline calibration with Common Astronomy Software Application \citep[CASA;][]{McMullin2007a} version 5.4.0. The bandpass and amplitude calibrations were performed with the quasar J0423$-$0120, and the phase calibration was conducted with the quasar J0510$-$1800. We performed self-calibration on the phase with the continuum emission. The self-calibration process was iterated four times with decreasing solution intervals of \texttt{solint = inf}, \texttt{inf}, \texttt{90s} and \texttt{30s} from the first to last iterations. All scans were averaged in the first round, but not in the subsequent iterations. Then, the solutions were applied to the \ce{H2CO} line data.

In addition, the self-calibrated \ce{HCO+} ($3\mbox{--}2$) line data (267.557633 GHz) presented by \cite{Yen2019a} were revisited for comparison. The details of the observations and data calibration process, including the self-calibration, were described in \cite{Yen2019a}. For brief summary, the data were calibrated with the pipeline calibration with CASA version 4.7.0. Self-calibration on the phase was performed for the continuum data, and then the solution was applied to the line data.

We performed imaging using the \texttt{tclean} task in CASA version 6.5.2. Briggs weighting with a robust parameter of 0.5 was applied to both \ch{H_2CO} and \ch{HCO^+} data. The resulting angular resolutions of the \ch{H_2CO} and \ch{HCO^+} image cubes are $0\farcs325 \times 0\farcs275~(-1.8^\circ)$ and $0\farcs095\times0\farcs088~(20.2^\circ)$, and their rms noise levels are $1.68~\mjypbm$ and $2.03~\mjypbm$, respectively. The channel width of the \ce{H2CO} and \ce{HCO+} images was set to $0.163~\kmps$, which was the finest spacing allowed for the \ce{H2CO} data.

\subsection{Observational Results} \label{subsec:obs_result}

% ------------ Figure -------------
\begin{figure*}
    \centering
    \includegraphics[width=\linewidth]{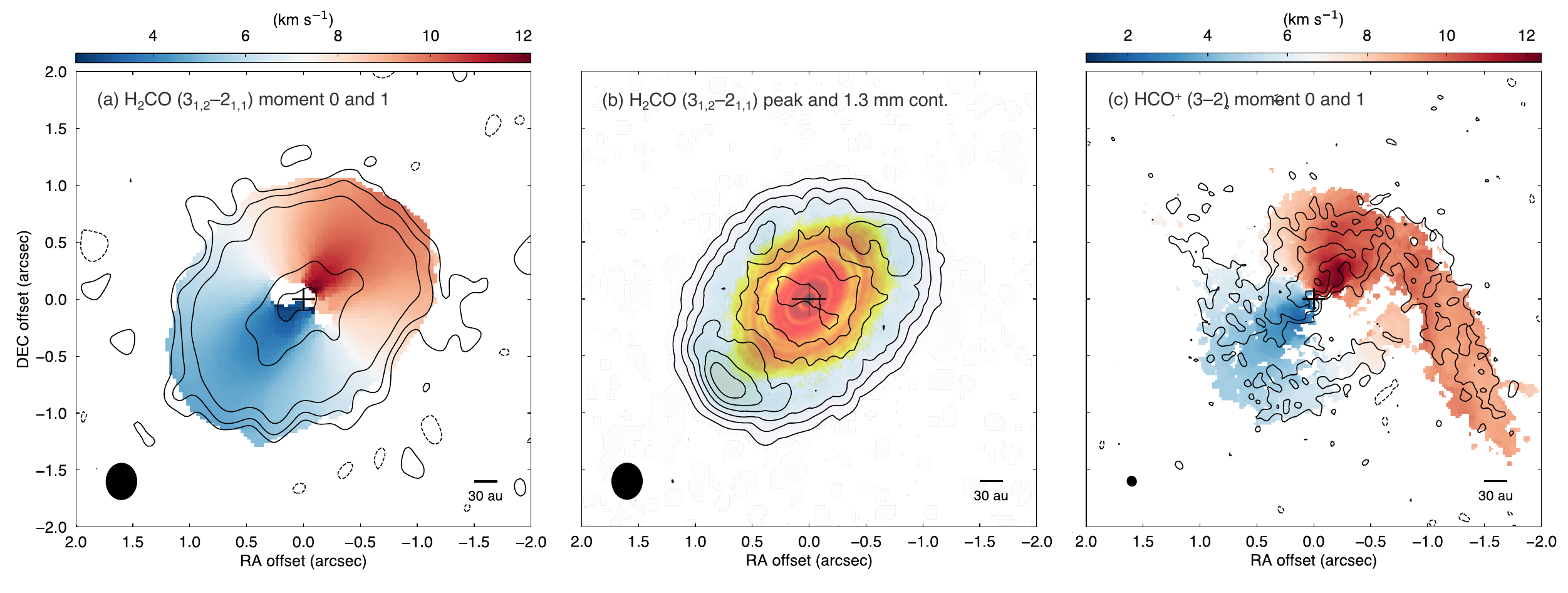}
    \caption{(a) Moment 0 and 1 maps of the \ce{H_2CO} ($3_{1,2}\mbox{--}2_{1,1}$) emission, shown with contours and color, respectively. Contour levels are $-3\sigma$, $3\sigma$, $6\sigma$, and $12\sigma$, where $\sigma= 2.37 ~\mjypbm~\kmps$. (b) The 1.3 mm dust continuum emission (orange color) overlaid with the peak intensity map of the \ce{H_2CO} emission (contours and light blue color). Contours start from $5\sigma$ and increase in steps of $5\sigma$, where $\sigma= 1.68~\mjypbm$. (c) Moment 0 and 1 maps of the \ce{HCO^+} ($3\mbox{--}2$) emission, shown with contours and color, respectively. Contour levels are $-3\sigma$, $3\sigma$, $6\sigma$, and $12\sigma$, where $\sigma= 4.90 ~\mjypbm~\kmps$. The filled ellipses at the bottom left corners denote the beam size of each image.}
    \label{fig:maps}
\end{figure*}

% Channel maps
\begin{figure*}
    \centering
    \includegraphics[width=\linewidth]{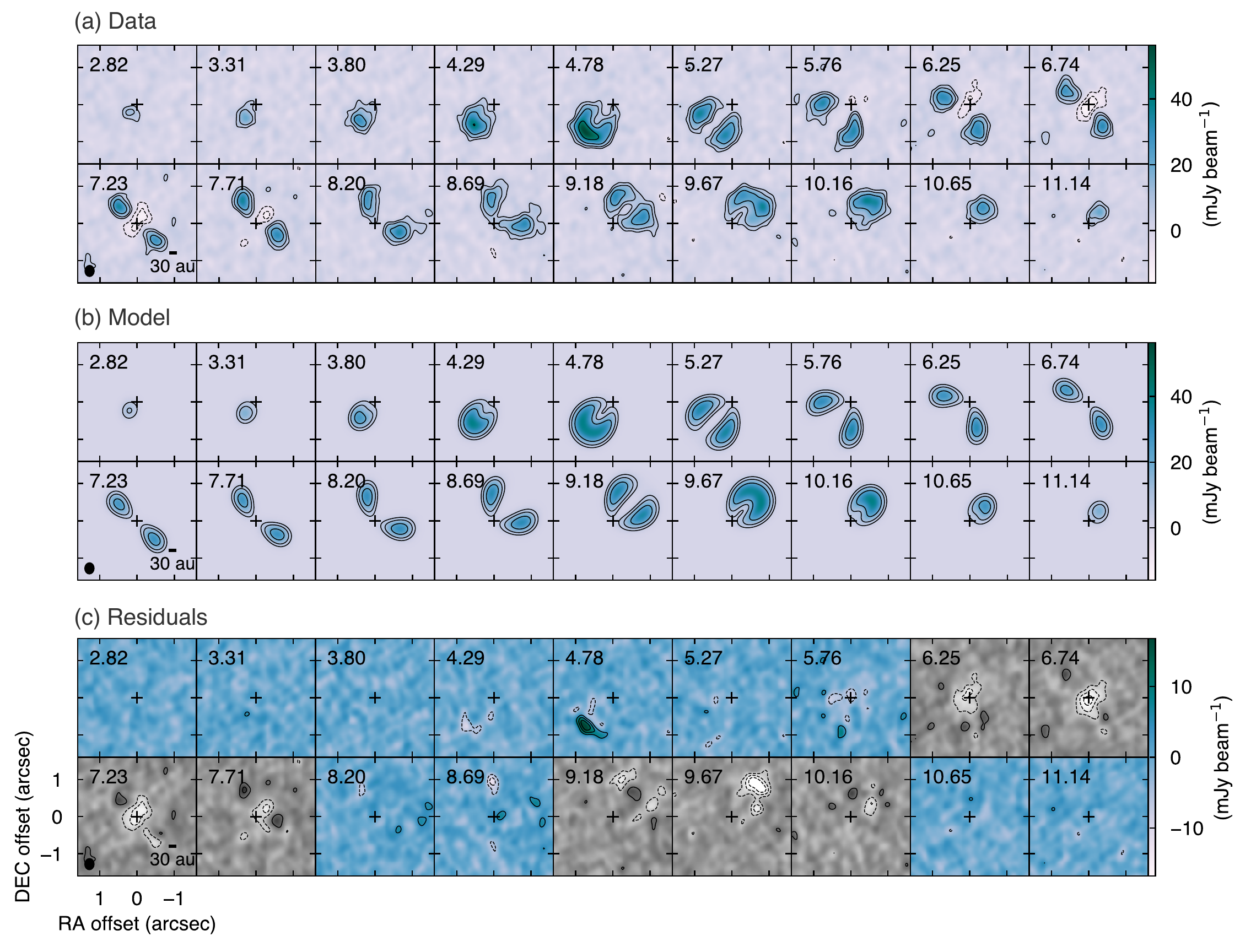}
    \caption{Velocity channel maps of the \ce{H_2CO} ($3_{1,2}\mbox{--}2_{1,1}$) data, the best-fit model, and residuals after subtracting the best-fit model from the data. The label at the top left corner of each panel denotes the LSR velocity. The velocity channels are presented in steps of three times original channel spacing. Contour levels are $-6\sigma$, $-3\sigma$, $3\sigma$, $6\sigma$, $12\sigma$ and $24\sigma$, where $\sigma= 1.68 ~\mjypbm$. Dashed and solid contours indicate negative and positive values, respectively. The filled ellipses at the bottom left corners denote the beam size. Gray scales indicate the channels that are excluded from the parametric model fitting because of absorption by foreground envelope gas or contamination of emission from the main streamer.}
    \label{fig:channel_fitres}
\end{figure*}
% ----------------------------------

Figure \ref{fig:maps}a presents moment 0 and 1 maps of the \ch{H_2CO} emission. The \ch{H_2CO} emission shows a clear velocity gradient in the direction of the disk major axis \citep[P.A. of $138^\circ$;][]{ALMAPartnership2015b}. This velocity gradient and the disk-like morphology of the moment 0 map suggest that the \ce{H_2CO} emission primarily traces a Keplerian disk. In a previous study, the moment 1 map of the \ce{H_2CO} emission is indeed well fitted by a Keplerian disk model with a stellar mass of $M_\ast=2.1~\Msun$ \citep{Garufi2022a}. The moment 0 map appears to be a ring with the intensity peak around a radius of $\rtsim100~\au$. This ring shape arises from the subtraction of the optically thick dust continuum emission behind the \ch{H_2CO} emission.

The velocity channel maps of the \ch{H_2CO} emission are presented in Figure \ref{fig:channel_fitres}a. The channel maps show the typical butterfly pattern of a Keplerian disk. The \ch{H_2CO} emission appears to follow a single isovelocity curve in each velocity channel. When the emitting surface is located at a high altitude from the disk midplane because of the high optical depth or the vertical stratification of molecules, the emission follows two isovelocity curves tracing the upper and lower surfaces of the disk in each velocity channel \citep{Pinte2018a}. Hence, the \ch{H_2CO} emission is expected to arise predominantly from layers near the disk midplane. A more quantitative analysis to evaluate the emission height is presented in Section \ref{sec:emission_height}. The \ch{H_2CO} emission exhibits negative intensities around the disk center in velocity channels near the systemic velocity of $7.24~\kmps$, suggesting absorption by the foreground gas colder than the optically thick dusty disk. Nevertheless, the absorption is weak such that the \ch{H_2CO} emission from outer radii of the disk is clearly detected in these velocity channels.

Figure \ref{fig:maps}b shows a comparison between spatial distributions of the 1.3 mm dust continuum emission at a high angular resolution, presented by \cite{ALMAPartnership2015b}, and the peak intensity map of the \ce{H_2CO} emission. The \ce{H_2CO} emission exhibits the highest signal-to-noise ratio (S/N) just outside the dusty disk exhibiting substructures, and extends up to a radius of $\rtsim200~\au$. Therefore, the \ce{H_2CO} emission mainly traces outer regions compared to the dusty disk.

For comparison, the \ch{HCO^+} emission is also presented in Figure \ref{fig:maps}c. The emission exhibits a clear velocity gradient in the direction of the disk major axis, which likely traces the rotational motion of the disk. In addition, faint blueshifted emission and more prominent redshifted emission are elongated to the northeast and southwest from the disk, respectively. As discussed in detail in \cite{Yen2019a}, velocity structures of these two components are consistent with infalling motion and they likely trace infalling gas flows toward the disk. The \ch{HCO^+} emission appears to be missing in the southwest of the protostar, which is due to absorption by the foreground envelope \citep{Yen2019a}.

The absence of clear infalling streamers in the \ce{H_2CO} emission suggests that it traces the Keplerian disk with little contamination of the infalling envelope. Motivated by these observational results, the \ce{H_2CO} emission is used as a disk tracer and further analyzed to measure turbulence from the nonthermal line broadening in the disk.

\section{Parametric Modeling} \label{sec:modeling}

The microscopic nonthermal gas motion can be derived from the line widths of the observed spectra. The total line width of the observed \ch{H_2CO} spectrum at a given spatial coordinate is, however, dominated by Keplerian shear motion due to the limited spatial resolution of the data and small disk size. Hence, we employed a parametric model fitting approach to measure the nonthermal velocity dispersion taking into account the Keplerian shear motion. In this method, we fitted velocity channel maps of the \ch{H_2CO} emission with a parametric disk model to constrain the physical parameters that determine the velocity structures of the emission. The local line width, including both thermal and nonthermal broadening, was parameterized as a power-law function of radius in the model assuming no vertical variations. The temperature structure of the HL Tau disk has been well characterized in previous studies \citep{Okuzumi2016a, Yen2019a}. Hence, we evaluated the thermal component using the temperature profiles constrained in previous work. Then, the nonthermal velocity dispersion was derived by subtracting the thermal component from the local line width that was determined by the parametric model fitting.

\subsection{Description of Fiducial Model}
\label{subsec:description_fiducial_model}

% ------------ Figure -------------
\begin{figure*}
    \centering
    \includegraphics[width=\linewidth]{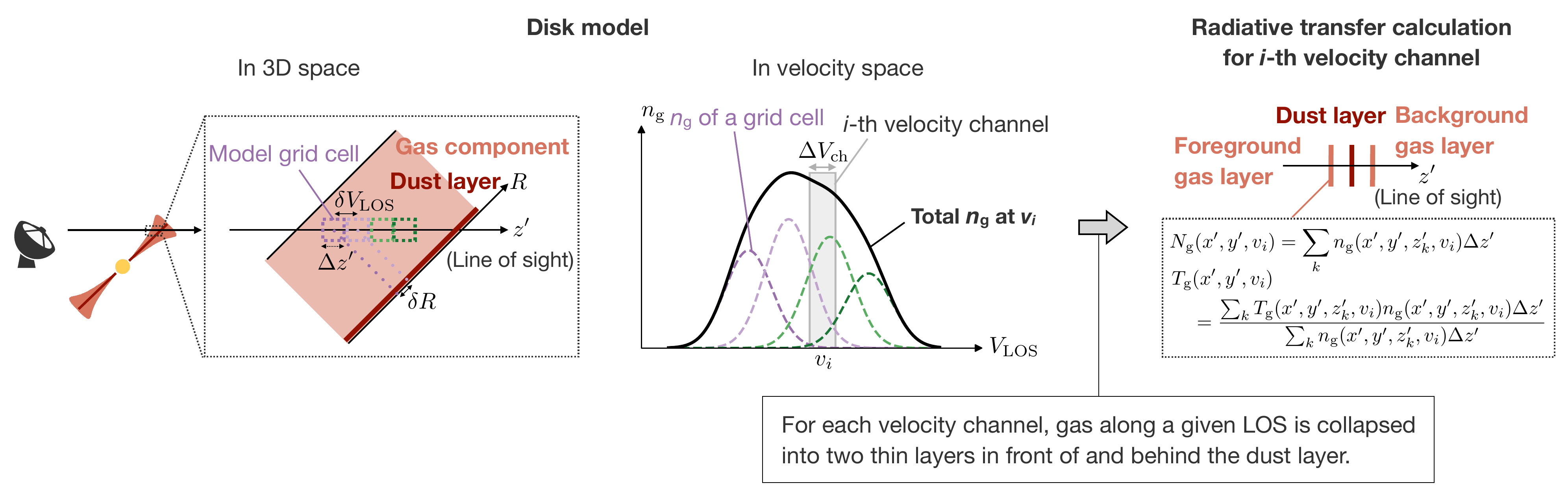}
    \caption{Schematic overview of the channel-based three-layer approximation adopted in the radiative transfer calculation in this work. The illustration represents a single line of sight (LOS) corresponding to a position $(x',y')$ on the plane of the sky.}
    \label{fig:concept_rt}
\end{figure*}
% ----------------------------------

% --------- Table ---------
\begin{deluxetable}{lllcc}
%\digitalasset
%\tablewidth{\columnwidth}
\tablecaption{Parameters of the fiducial model for the line fitting \label{tab:params}}
\tablehead{
\colhead{Parameter} & \colhead{Unit} & \colhead{Description of parameter} & \colhead{Best-fit value} & \colhead{Prior range}
}
\startdata
\multicolumn{5}{c}{Free} \\
$T_\mathrm{g,0}$ & K & Gas temperature at a radius of $100~\au$ & $ 23.0 \pm 0.3$ & (10, 35) \\
$q_\mathrm{g}$ & - & Power-law index of the gas temperature profile & $0.61 \pm 0.02 $ & (0.3, 1.0) \\
$\log_{10} N_\mathrm{g,c}$ & $\mathrm{cm}^{-2}$ & Gas surface density at the characteristic radius & $15.6\pm 0.1$ & (14, 18)  \\
$R_\mathrm{g,c}$ & au & Gas characteristic radius &  $92 \pm 4$ & (50, 150) \\
$\gamma_\mathrm{g}$ & - & Power-law index of the gas surface density profile & $-0.7 \pm 0.1$ & (-4, 1) \\
$i$ & $^\circ$ & Inclination angle & $45.4 \pm 0.4$& (42, 50) \\
$\theta_\mathrm{PA}$ & $^\circ$ & Position angle & $135.5\pm0.2$ & (130, 150) \\
$M_\ast$ & $\Msun$ & Central stellar mass & $2.34 \pm 0.02$ & (1.8, 2.5) \\
$\vsys$ & $\kmps$ & Systemic velocity & $7.237 \pm 0.006$ & (7.0, 7.4) \\
$\delta x_0$ & au & $x$-coordinate offset of the disk center relative to the map center &  $0.6 \pm 0.3$ & (-10, 10) \\
$\delta y_0$ & au & $y$-coordinate offset of the disk center relative to the map center & $-3.5 \pm 0.4$ & (-10, 10) \\
$\Delta V_0$ & $\kmps$ & Local line width at a radius of $100~\au$ & $0.181 \pm 0.008$ & (0.0, 0.5) \\
$l$ & - & Power-law index of the line width profile &  $-0.3 \pm 0.1$ & (0.0, 0.5) \\ \hline
\multicolumn{5}{c}{Fixed$^1$} \\
$T_\mathrm{d,0}$ & K & Dust temperature at a radius of $100~\au$ & 22.5 & - \\
$q_\mathrm{d}$ & - & Power-law index of the dust temperature profile & 0.57 & - \\
$\log_{10} \tau_\mathrm{d,c}$ & - & Dust optical depth at the dust characteristic radius & 0.317 & - \\
$R_\mathrm{d,c}$ & au & Dust characteristic radius &  $84.2$ & - \\
$\gamma_\mathrm{d}$ & - & Power-law index of the dust optical depth profile & $-0.2$ & - \\
\enddata
\tablecomments{$^1$We adopted $T_\mathrm{d,0}$ and $q_\mathrm{d}$ from \cite{Okuzumi2016a}, and $R_\mathrm{d,c}$ and $\gamma_\mathrm{d}$ from \cite{Kwon2015a} with correction for an updated distance. We determined $\log_{10} \tau_\mathrm{d,c}$ by fitting the continuum emission obtained from line-free channels of the \ch{H_2CO} data, using $T_\mathrm{d}$, $q_\mathrm{d}$, $R_\mathrm{d,c}$ and $\gamma_\mathrm{d}$ taken from the literature.}
\end{deluxetable}
% --------------------------

In this subsection, we describe the fiducial disk model that is used to fit the observational data. We employ the standard viscous disk model \citep[][]{Shakura1973a}. The model is constructed in three dimensions with a set of formulae given in cylindrical coordinates $(R, z, \phi)$. The surface density of the \ch{H_2CO} molecule is expressed as follows assuming a constant molecular abundance throughout the disk:
\begin{eqnarray}
    N_{\mathrm{g}} (R) = N_\mathrm{g,c} \left( \frac{R}{R_\mathrm{g,c}} \right)^{-\gamma_\mathrm{g}} \exp\left\{-  \left(\frac{R}{R_\mathrm{g,c}} \right)^{2-\gamma_\mathrm{g}} \right\}.
    \label{eq:gas_surface_density}
\end{eqnarray}
The vertical structure of the gas is assumed to follow the hydrostatic equilibrium:
\begin{eqnarray}
    n_{\mathrm{g}} (R) = \frac{N_\mathrm{g}(R)}{\sqrt{2\pi}H} \exp\left(- \frac{z^2}{2H^2} \right).
    \label{eq:gas_volume_density}
\end{eqnarray}
Here, $H\equiv \cs / \Omega$ is the gas pressure scale height, where $\cs$ is the isothermal sound speed and $\Omega$ is the angular velocity. The isothermal sound speed is defined as $\cs = \sqrt{\kb T / \mu \mH}$ with the Boltzmann constant $\kb$, the gas temperature $T$, the mean molecular weight $\mu = 2.37$ and the weight of the hydrogen atom $\mH$. The gas pressure scale height is calculated with the gas temperature and rotational velocity field described below. As found in Equation (\ref{eq:gas_volume_density}), \ch{H_2CO} molecules are assumed to be distributed near the disk midplane, since the velocity channel maps suggest the emission arises from layers close to the disk midplane. However, we note that contributions from higher altitudes are also accounted for in the model, as the modeled disk has a vertical thickness. Another model with elevated molecular layers, whose the \ch{H_2CO} density peaks at a finite vertical height above the midplane, is discussed separately in Section \ref{sec:emission_height}. The gas kinetic temperature is assumed to follow a power-law function:
\begin{eqnarray}
    T_{\mathrm{g}} (R) = T_\mathrm{g,0} \left( \frac{R}{100~\au} \right)^{-q_\mathrm{g}}.
    \label{eq:gas_temp}
\end{eqnarray}
The gas disk is assumed to be vertically isothermal. Then, the gas kinetic temperature $T_\mathrm{g}$ is adopted for the excitation temperature $\tex$ of the \ch{H2CO} emission assuming the local thermal equilibrium (LTE).

The model includes a geometrically thin dust layer to take into account the effect of subtraction of the optically thick dust continuum. We model the dust layer separately from the gaseous component to avoid assumptions about the gas-to-dust mass ratio and molecular abundance, both of which are poorly constrained. The radial profile of the optical depth of the dust layer is described as
\begin{eqnarray}
    \tau_{\nu, \mathrm{d}} = \tau_{\mathrm{d,c}} \left( \frac{R}{R_\mathrm{d,c}} \right)^{-\gamma_\mathrm{d}} \exp\left\{-  \left(\frac{R}{R_\mathrm{d,c}} \right)^{2-\gamma_\mathrm{d}} \right\}.
\end{eqnarray}
This form is motivated by the viscous disk model and an assumption of a constant dust absorption coefficient across the disk, leading to $\tau_{\nu, \mathrm{d}} \propto \Sigma_\mathrm{d}$. The dust temperature profile is assumed to be the same form as that for gas but is modeled independently with $T_\mathrm{d,0}$ and $q_\mathrm{d}$. For simplicity, the dust layer is assumed to have smooth density and temperature profiles, ignoring the ring and gap structures that are known in HL Tau, which is sufficient for the current purpose of including the effect of the subtraction of the dust continuum emission at the current spatial resolution.

The velocity field of the gas is assumed to follow Keplerian motion including the vertical height:
\begin{eqnarray}
    \frac{v_\phi^2}{R} = \frac{GM_\ast R}{(R^2 + z^2)^{3/2}},
    \label{eq:vkep}
\end{eqnarray}
where $M_\ast$ is the protostellar mass, $G$ is the gravitational constant, and $v_\phi$ is the rotational velocity. No systemic motions in the radial or vertical directions are adopted. The line-of-sight (LOS) velocity is calculated based on the projection of the rotational velocity onto the plane of the sky (POS) as
\begin{eqnarray}
    \vlos = v_\phi \cos \phi \sin i,
    \label{eq:vlos}
\end{eqnarray}
where $i$ is the inclination angle of the disk. Then, rotation by the position angle (PA; $\theta_\mathrm{PA}$) and a shift of the disk center from the image center $(\delta x_0, \delta y_0)$ are applied on POS coordinates.

The line profile function at each three dimensional position is calculated as
\begin{eqnarray}
    \psi (v) = \frac{1}{\sqrt{\pi}\Delta V} \exp\left\{{-\frac{(\vlos - v )^2}{\Delta V^2}} \right\},
    \label{eq:lnprof}
\end{eqnarray}
where $\Delta V$ is the local line width and $\mathrm{FWHM} = 2 \sqrt{\ln 2} \Delta V$, considering the Doppler line broadening. The local line width is modeled with a power-law function of radius
\begin{eqnarray}
    \Delta V (R) = \Delta V_0 \left(\frac{R}{100~\au}\right)^{l}.
    \label{eq:dv_prof}
\end{eqnarray}
This local line width accounts for all sources of line broadening except for the Keplerian shearing and is independent of the assumed gas temperature in the model. This allows us to later assess uncertainties in the gas temperature and thermal line broadening that cannot be fully considered in the simple parametric model. All the model parameters are summarized in Table \ref{tab:params}.

\subsection{Radiative Transfer with Channel-based Three-layer Approximation}

Previous measurements of turbulent line broadening in protoplanetary disks have commonly employed full radiative transfer forward modeling \citep[e.g.,][]{Flaherty2020a}, or simplified intensity distribution models that approximate geometrically thin emitting layers \citep{Paneque-Carreno2024a}. The former approach is physically rigorous and naturally accounts for the effects of disk geometry and optical depth. However, full radiative transfer calculations are computationally expensive, making it difficult to efficiently explore a large parameter space. The latter method without the radiative transfer is more efficient but does not fully account for the effects of disk geometry and optical depth. In particular, in an inclined disk with finite thickness, gas located at different radii can lie along a common LOS, leading to velocity differences that contribute to line broadening. This velocity shear induced by the finite thickness of the disk is more significant at smaller radii \citep{Paneque-Carreno2024a}, and needs to be taken into account in the current case of HL Tau given its small disk size.

In this work, we compute the radiative transfer to account for the effect of the finite thickness of the disk, but with the channel-based three-layer approximation introduced below, which significantly reduces the computational cost and allows exploring a large parameter space. Using this approximation, we simultaneously constrain the local line width and the model parameters listed in Table \ref{tab:params} to reduce systemic biases in the derived line width that could arise from fixing other model parameters.

The concept of the channel-based three-layer approximation is summarized in Figure \ref{fig:concept_rt}. The emission in a given velocity channel is only contributed by gas whose LOS velocities fall within the channel bin. Therefore, we first calculate the LOS velocity distribution of the gas in the three dimensional disk model, following the description in Section \ref{subsec:description_fiducial_model}. The gas within each model grid cell follows a Gaussian line profile centered on the projected Keplerian velocity, as given by Equation (\ref{eq:vlos}) and (\ref{eq:lnprof}). Hence, gas within a single grid cell contributes to multiple velocity channels, as illustrated in the middle of Figure \ref{fig:concept_rt}. For each velocity channel, we then collapse the gas associated with that channel into two effective thin layers in front of and behind the dust layer, respectively. The column density of each gas layer is computed by integrating the gas densities within the velocity channel along the LOS, while the gas temperature of each layer is computed as the density-weighted mean temperature. We separately treat the foreground and background gas layers to account for radiative interactions among them. This treatment is required because the foreground and background gas can share the same LOS velocity,  particularly along the disk major and minor axes on the POS \citep[e.g.,][]{Pinte2018a}.

This approximation reduces the radiative transfer problem for each velocity channel to a three-layer system consisting of a foreground gas layer, a dust layer, and a background gas layer, as depicted in the right of Figure \ref{fig:concept_rt}. As a result, the radiative transfer equation can be solved analytically, which significantly reduces the computational cost compared to full ray tracing through all gas cells. At the same time, the approximation naturally retains the line broadening caused by the finite thickness of the disk because the gas is first assigned to velocity channels according to its LOS velocity before being collapsed into effective layers. Consequently, gas at different heights along the same LOS is assigned to different velocity channels if the velocity shear is sufficiently large. Since the collapse is performed independently for each velocity channel, the velocity shear associated with the finite disk thickness is preserved.

The optical depth of the molecular line of the gas layers is calculated assuming LTE. Under the above assumptions, the radiative transfer equation at each frequency and POS coordinate is written as
\begin{align}
    I_\mathrm{\nu, line} (x',y') = & B_\nu (T_\mathrm{bg}) e^{-\tau_{\nu,\mathrm{gb}} - \tau_\mathrm{\nu,d} - \tau_\mathrm{\nu,gf}} \nonumber \\
    & + B_\nu (T_\mathrm{gb}) \left( 1 - e^{-\tau_\mathrm{\nu,gb}}\right) e^{- \tau_\mathrm{\nu,d} - \tau_\mathrm{\nu,gf}} \nonumber\\
    & + B_\nu (T_\mathrm{d}) \left( 1 - e^{-\tau_\mathrm{\nu,d}}\right) e^{- \tau_\mathrm{\nu,gf}} \nonumber\\
    & + B_\nu (T_\mathrm{gf}) \left( 1 - e^{-\tau_\mathrm{\nu,gf}}\right) \nonumber\\
    & - B_\nu (T_\mathrm{bg}).
    \label{eq:I_line}
\end{align}
Here, ($x'$, $y'$) is the POS coordinates. The subscripts, d, gf and gb, denote quantities associated with the dust layer, the foreground gas layer, and the background gas layer, respectively. The background temperature $T_\mathrm{bg}$ is assumed to be $2.73~\mathrm{K}$. The line emission after continuum subtraction is calculated with
\begin{align}
    \Delta I_\mathrm{\nu, line} = I_\mathrm{\nu, line} - I_\mathrm{\nu, dust},
\end{align}
where the dust continuum emission is given by
\begin{align}
   I_\mathrm{\nu, dust} = \left\{B_\nu (T_\mathrm{d}) - B_\nu (T_\mathrm{bg}) \right\} \left( 1 - e^{-\tau_\mathrm{\nu,d}}\right).
   \label{eq:I_dust}
\end{align}

The channel-based three-layer approximation should be regarded as an approximation in the radiative transfer calculation rather than a physical model in which the disk consists of only three discrete layers. In Appendix \ref{app:radmcmodel}, we compare models generated with this approximation and full radiative transfer calculations, and demonstrate that this approximation reproduces results of the full radiative transfer within the noise level of the current observational data.

In the signal processing for ALMA data, Hanning smoothing is applied in the lag domain prior to channel averaging, resulting in a spectral resolution slightly coarser than the channel spacing. To mimic this process, the model channel bin is first set to be three times finer than a FWHM of a smoothing function. Then, a Gaussian function with FWHM of $0.16~\kmps$, equivalent to the width of the smoothing function of the data, is convolved after solving the radiative transfer. Finally, the model channel bins are averaged down to the image channel spacing, and a Gaussian beam of the same size as that of the data is convolved.

\subsection{Fitting Data} \label{subsec:fitting}

We fitted the observational data with the model using the Markov Chain Monte Carlo (MCMC) method. The fitting was performed in two steps. First, we fitted the continuum emission obtained from line-free channels with a geometrically thin dusty layer model to determine the dust optical depth $\tau_\mathrm{d,c}$. The other parameters of the dusty disk are fixed based on previous studies. We adopted parameters for the dust surface density distribution $(R_\mathrm{c,d}, \gamma_\mathrm{d}) = (84.2~\au, -0.2)$ from \cite{Kwon2015a}, which used the same viscous disk model, with correction for the updated distance. We used the dust temperature profile $T_\mathrm{d}(R) = 22.5  (R/100~\au)^{-0.57}~\mathrm{K}$ derived in \cite{Okuzumi2016a}. Then, we fitted the channel maps of the \ch{H2CO} emission with the model described above using the derived $\tau_\mathrm{d,c}$ and the other parameters of the dust layer. We allowed the parameters of the gas surface density and temperature profiles to vary during the fitting to match the intensity of the model and observed image cubes. However, the derived gas surface density and temperature profile would not be very reliable because only a single rotational transition is used in the fitting. Therefore, uncertainties of the gas temperature and thermal line broadening were discussed separately.

Several velocity channels were excluded from the fitting procedure. The \ch{H2CO} emission exhibits negative intensities near the disk center in the velocity range of $5.92\mbox{--}7.71~\kmps$, as noted in Section \ref{subsec:obs_result}. These channels were therefore excluded, as they are likely affected by absorption from the foreground envelope. In addition, a careful inspection of the observed spectra indicates that channels between $8.86~\kmps$ and $10.32~\kmps$ are likely contaminated by emission associated with an infalling streamer, which is discussed in more detail in Section \ref{sec:fitres}. Hence, these velocity channels were also excluded from the fitting. We note that the line velocity dispersion can still be reasonably constrained despite the exclusion of some velocity channels, since the spatial extent of the emission in individual channel maps retains information on the line broadening \citep{Guilloteau2012a, Flaherty2015a, Flaherty2020a}.

The MCMC sampling was performed using the Python package \texttt{emcee} \citep{Foreman-Mackey2013a} with 26 walkers and 2500 steps, with the first 2000 steps discarded as burn-in. All fixed and best-fit parameters with their prior ranges in the fitting for the \ch{H_2CO} emission are summarized in Table \ref{tab:params}. The corner plot showing posterior distributions of the parameters is presented in Figure \ref{fig:corner} in Appendix \ref{app:supp_fig}.

\subsection{Measurement of Nonthermal Velocity Dispersion} \label{subsec:measure_vnth}

The nonthermal velocity dispersion was derived by subtracting the thermal component from the local line width $\Delta V$, which was constrained with the model fitting, using a relationship of
\begin{eqnarray}
    \Delta V = \sqrt{\vth^2 + \vnth^2},
    \label{eq:dv_def}
\end{eqnarray}
where $\vth$ and $\vnth$ are the thermal and nonthermal velocity dispersions, respectively. The thermal component is defined as
\begin{eqnarray}
    \vth = \sqrt{\frac{2 k T}{\mmol \mH}},
    \label{eq:vth}
\end{eqnarray}
where $\mmol$ is the molecular weight.

As noted earlier, the gas temperature inferred from the current model fitting is not well constrained, as only a single transition is used. Therefore, to estimate the thermal component, we adopted radial temperature profiles derived in previous observational studies. \cite{Okuzumi2016a} derived a dust midplane temperature profile of $T(R) = 22.5 (R/100~\au)^{-0.57}~\mathrm{K}$ using optically thick dust continuum emission. \cite{Yen2019a} performed an LTE analysis of the \ce{HCO^+} $J=3\mbox{--}2$ and $J=1\mbox{--}0$ transitions, finding the former line to be optically thick and deriving a kinetic temperature profile of $T_\mathrm{k}(R) = 28.1 (R/100~\au)^{-0.48}~\mathrm{K}$. Because the \ch{H_2CO} emission is expected to trace a height between the disk midplane and the more optically thick \ch{HCO^+} emitting layer, we adopted these dust- and \ch{HCO^+}-based temperature profiles as lower and upper bounds on the gas temperature, respectively, considering the typical disk thermal structure that monotonically increases with increasing heights. The upper limit temperature is also comparable to the temperature estimated by \cite{Guerra-Alvarado2024a} and \cite{Ueda2025a} based on the optically thick 0.45 mm dust continuum emission, which traces small dust grains that are well coupled to gas and thus could arise from a slightly elevated layer.

\cite{Carrasco-Gonzalez2019a} reported a high dust temperature profile of $T(R) = 47 (R/100)^{-0.5}$ based on multiwavelength dust continuum analyses. This temperature is substantially higher than other estimates of the dust temperature and also exceeds the gas temperature inferred from the \ch{HCO^+} emission. Such a discrepancy may arise from a high dust albedo assumed in their modeling. This dust temperature profile requires a vertical temperature inversion because the gas temperature estimated with the \ch{HCO^+} emission, which presumably traces a higher altitude, 
is lower. Furthermore, two dimensional temperature measurements of Class \II~disks using optically thick molecular lines suggest that disk temperatures are typically below $30~\kelv$ at radii of $R=100\mbox{--}200~\au$ and heights of $z/R \leq 0.1$ \citep{Law2021b}, corresponding to the region predominantly traced by the \ch{H_2CO} emission (see Section \ref{sec:emission_height} for a quantitative discussion of the \ch{H_2CO} emission height). This also supports a characteristic temperature range of $\tsim20\mbox{--}30~\kelv$ at $R=100~\au$. For these reasons, we do not adopt the temperature profile of \cite{Carrasco-Gonzalez2019a} when deriving the nonthermal velocity dispersion. The impact of adopting this higher temperature profile is discussed separately in Section \ref{subsec:uncertainties}.

\section{Fitting Results} \label{sec:fitres}

\subsection{Fit with Fiducial Model} \label{sec:fiducial_model}
% ------------ Figure -------------
\begin{figure}
    \centering
    \includegraphics[width=\linewidth]{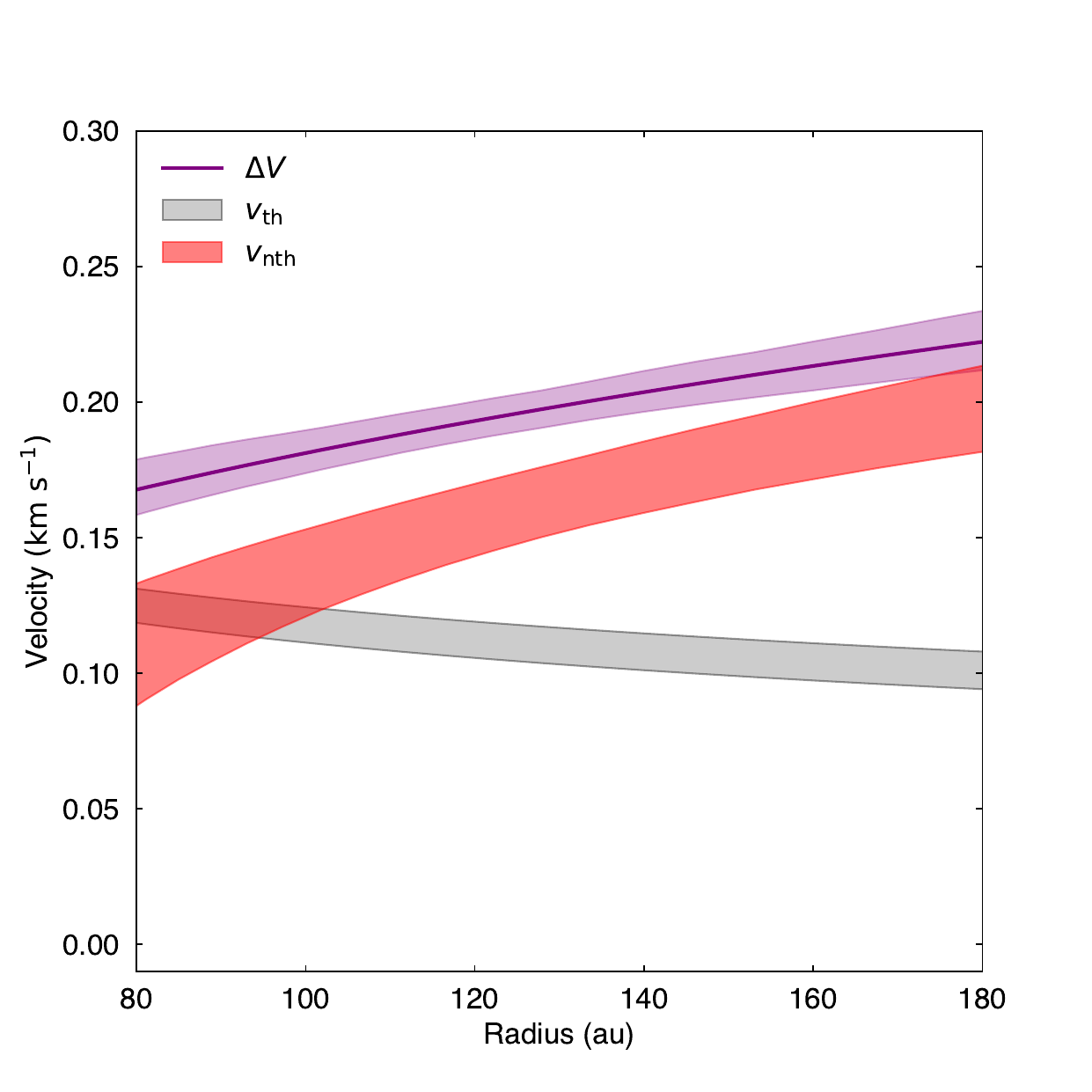}
    \caption{Radial profiles of the best-fit local line width, the thermal velocity component estimated using temperature distributions derived in \cite{Okuzumi2016a} and \cite{Yen2019a}, and the nonthermal velocity dispersion obtained by subtracting the thermal component from the best-fit local line width. The color shaded regions indicate the $1\sigma$ uncertainty of $\Delta V$, derived from the posterior distributions of $\Delta V_0$ and $l$, the error of the thermal velocity given the temperature range, and the uncertainty of the nonthermal velocity dispersion, derived from above two uncertainties.}
    \label{fig:dvprof_fitres}
\end{figure}
% ----------------------------------

% ------------ Figure -------------
\begin{figure*}
    \centering
    \includegraphics[width=\linewidth]{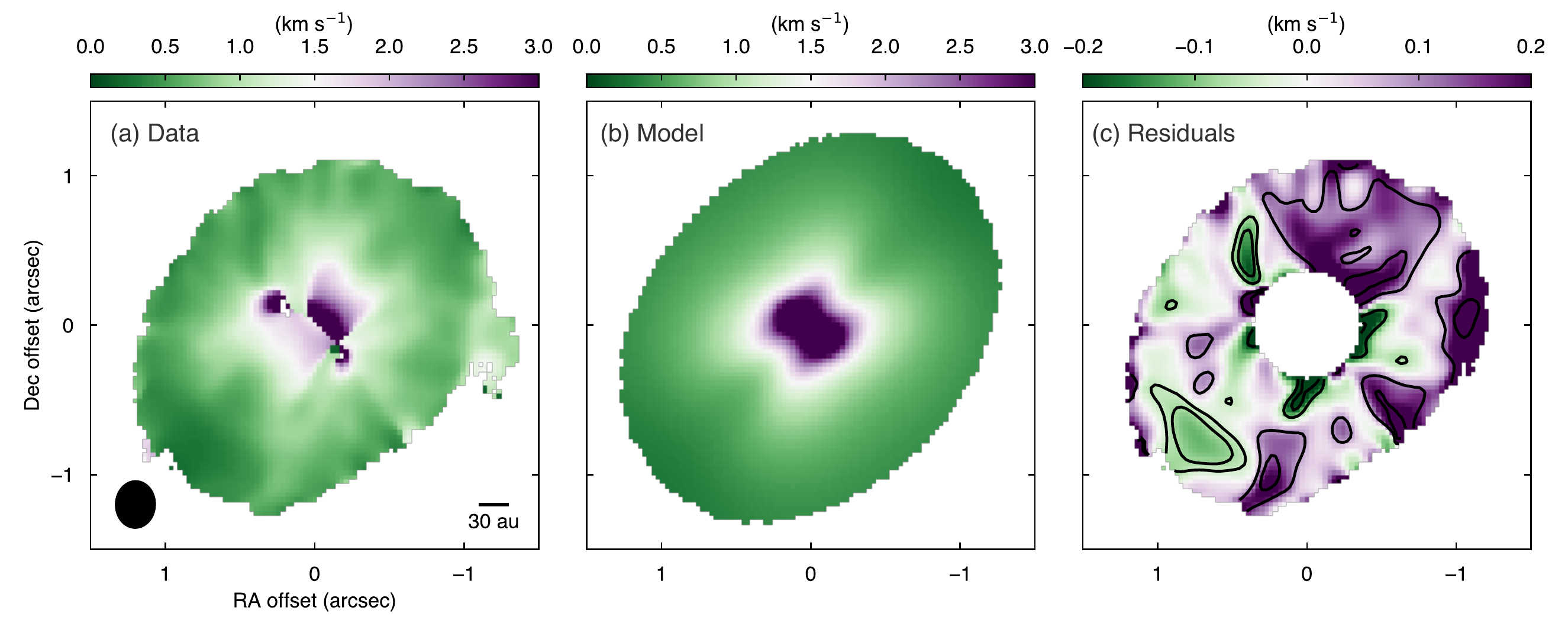}
    \caption{(a, b) Total line width maps of the data and best-fit model, including Keplerian shear motions. (c) The residual map after subtracting the total line widths of the best-fit model from those of the data. Contours indicate $\pm3\sigma$ and $\pm6\sigma$, where $\sigma$ is the fitting uncertainty derived in the Gaussian fitting. The central region, which shows large residuals due to absorption by foreground gas against the optically thick dust continuum in the data, is masked for better visualization.}
    \label{fig:dvmaps_fitres}
\end{figure*}
% ----------------------------------

% ------------ Figure -------------
\begin{figure}
    \centering
    \includegraphics[width=\linewidth]{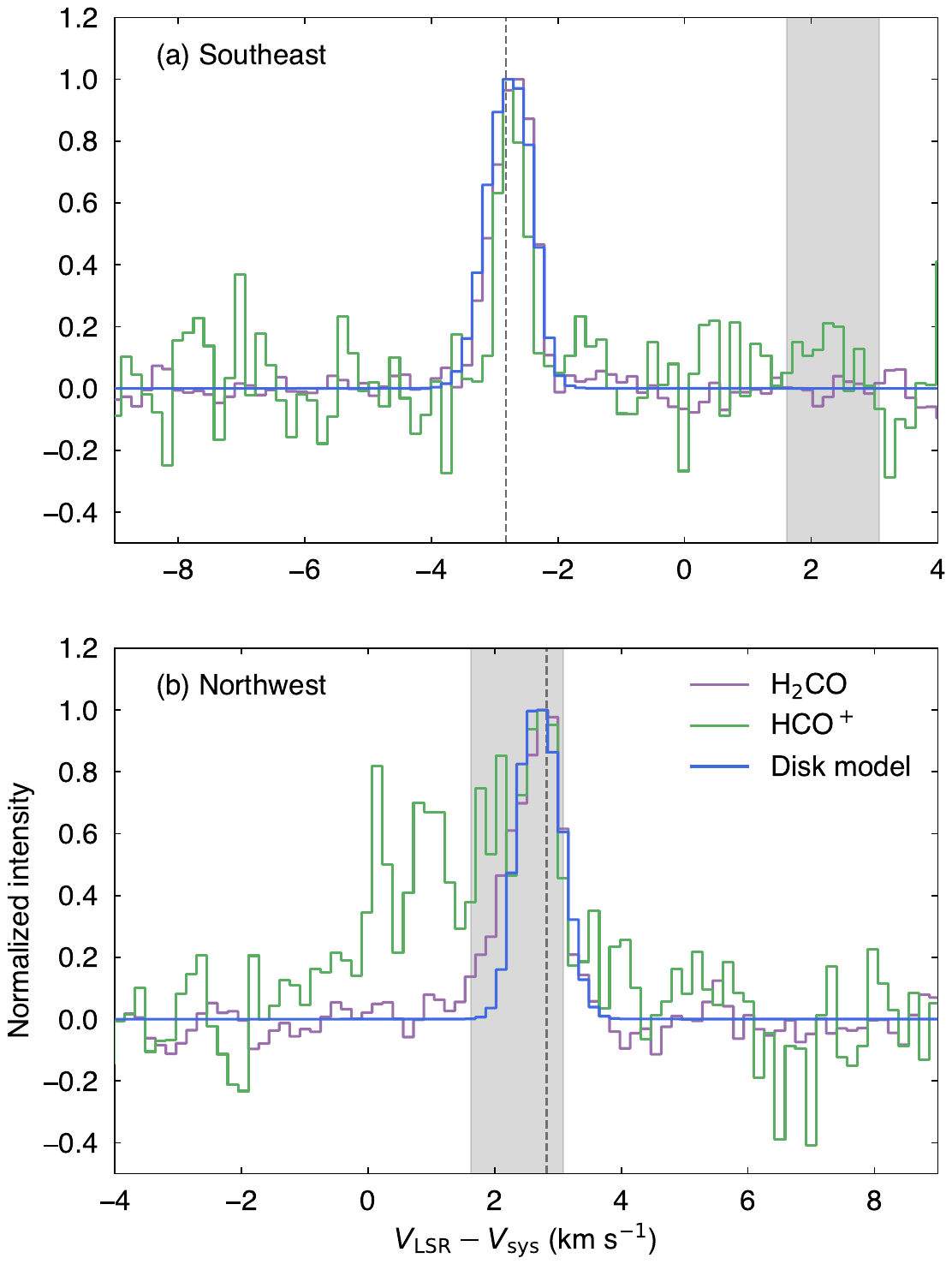}
    \caption{Spectra of the \ce{H2CO} emission, \ce{HCO^+} emission and the best-fit model measured at positions $0\farcs9$ away toward (a) southeast and (b) northwest from the protostellar position along the disk major axis. The gray shaded areas denote the velocity range that would be contaminated by the streamer component, and thus excluded from the model fitting. The vertical dashed lines indicate the projected Keplerian velocity at this radius. The systemic velocity is $\vsys = 7.24~\kmps$. The narrower line width of the \ch{HCO^+} emission in panel a arises from reduced Keplerian shear due to the smaller beam size.}
    \label{fig:spec_comp}
\end{figure}
% ----------------------------------

% ------------ Figure -------------
\begin{figure}
    \centering
    \includegraphics[width=\linewidth]{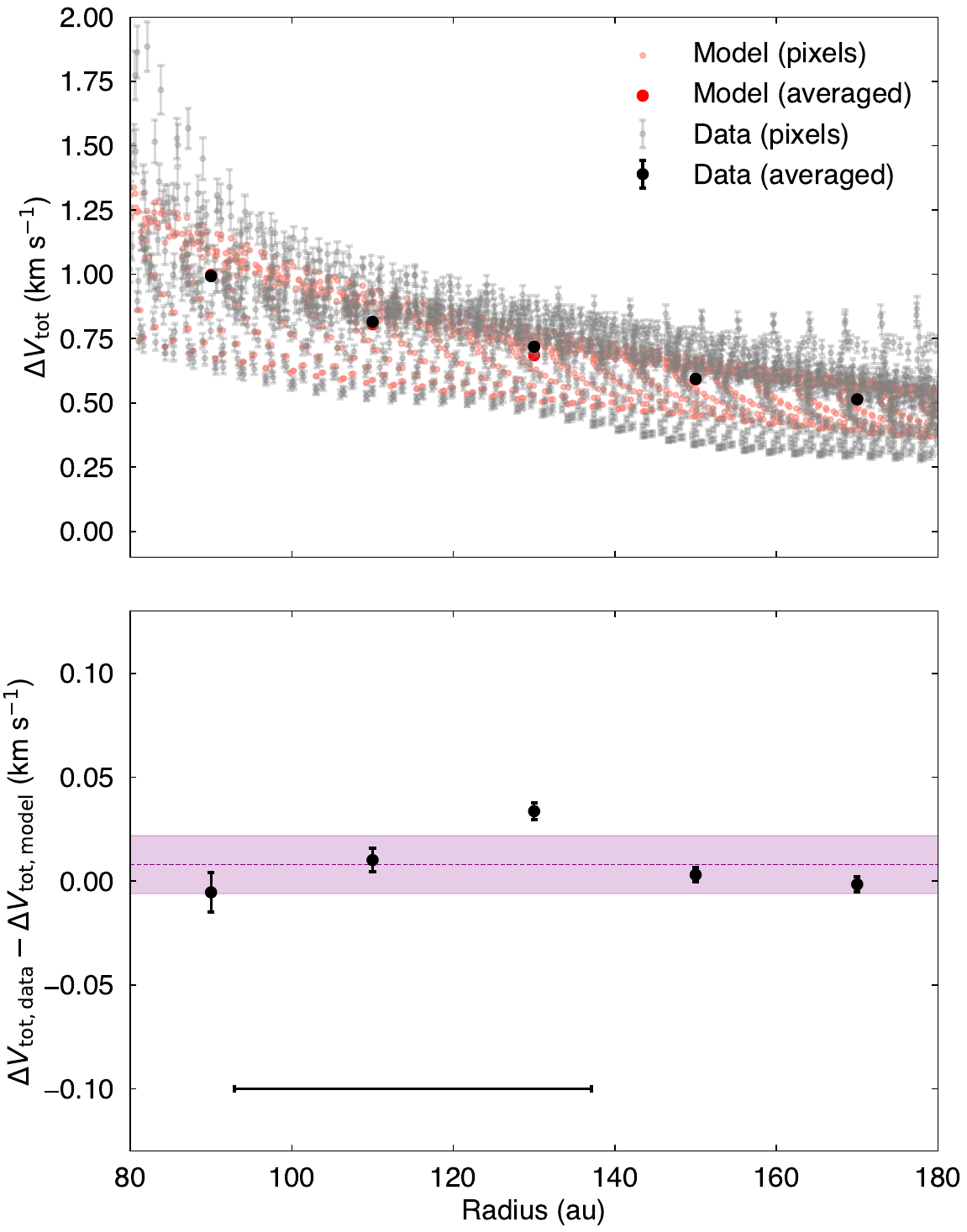}
    \caption{Radial profiles of the total line width measured with the data and best-fit model before and after azimuthal averaging (top), and residuals after azimuthal averaging (bottom). A half of the disk on the southeast side is used for these plots to avoid contamination of the streamer component. The horizontal bar indicates a mean beam size.}
    \label{fig:dvprof_comp}
\end{figure}
% ----------------------------------

% ------------ Figure -------------
\begin{figure}
    \centering
    \includegraphics[width=\linewidth]{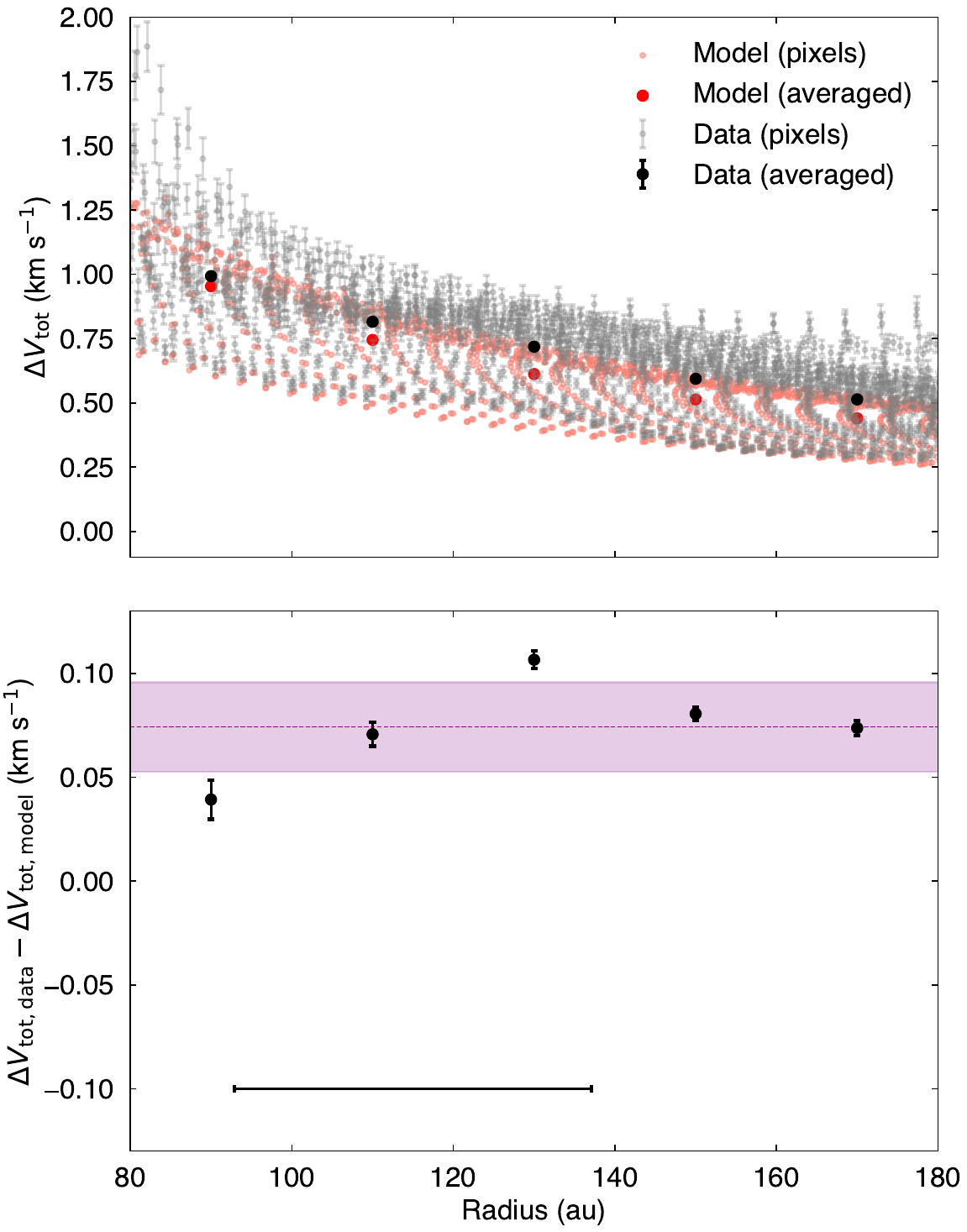}
    \caption{Same as Figure \ref{fig:dvprof_comp} but for the model only with the thermal line broadening.}
    \label{fig:dvprof_secfit_comp}
\end{figure}
% ----------------------------------

The velocity channel maps of the \ce{H_2CO} data and the best-fit fiducial model are shown in Figure \ref{fig:channel_fitres}a and b. The Keplerian disk model reproduces overall velocity structures of the \ce{H_2CO} emission well, which confirms that the \ce{H_2CO} emission primarily traces the Keplerian disk. Figure \ref{fig:dvprof_fitres} presents the radial profile of the local line width of the best-fit model over the radius, where the \ce{H_2CO} emission has S/N higher than $\rtsim12$. The purple shaded region indicates the $1\sigma$ uncertainty of $\Delta V$, derived from the posterior distributions of $\Delta V_0$ and $l$ obtained through the MCMC fitting. The local line width slightly increases with increasing radius. Figure \ref{fig:dvprof_fitres} also shows the thermal velocity dispersion estimated with the dust- and \ch{HCO^+}-based temperature profiles, and the resulting nonthermal velocity dispersion. The range of the nonthermal velocity dispersion is derived from the range of the thermal velocity and the $1\sigma$ uncertainty of $\Delta V$. The derived nonthermal velocity dispersion increases from $\rtsim0.10~\kmps$ to $\rtsim0.20~\kmps$ with increasing radius from $R\rtsim80~\au$ to $180~\au$. These values are significant compared to the uncertainties arising from the statistical errors of $\Delta V$ and uncertainties in the gas temperature. We note that the radial dependence would represent a global trend, but should not be considered to be an exact radial profile because of the spatial resolution of the data and limited flexibility of the parametric modeling approach. The average nonthermal velocity dispersion is $\rtsim0.15~\kmps$, corresponding to $0.6~\cs$, assuming $T=20~\mathrm{K}$ and $\mu=2.37$.

To further examine how well the best-fit model represents the line width of the observed data, we fitted the data and model spectra using a Gaussian line profile function pixel by pixel, and compared the total line width distributions. Figure \ref{fig:dvmaps_fitres} presents the derived total line width maps. The total line widths include contributions from Keplerian shear, and are therefore larger than the derived local line width. The total line width of the data is much smaller than that of the model near the disk center. This difference arises because the observed emission exhibits negative intensities near the disk center at low velocities due to absorption by the foreground gas with the optically thick dust continuum behind, whereas the model retains weak emission at these velocities. As a result, the observed spectra appear narrower than the model spectra.

Figure \ref{fig:dvmaps_fitres}c shows residuals after subtracting the total line widths of the best-fit model from those of the data. The central part showing a large deviation due to the above effect is masked for better visualization. Large positive residuals are present on the northwest side, i.e., the main-streamer side. These large line widths on the streamer side would likely be attributed to contamination of the streamer velocity component, which is excluded from the fitting. Figure \ref{fig:spec_comp} shows representative spectra, which are measured at $R=0\farcs9$ (or $R\tsim130~\au$) on the southeast and northwest sides of the protostar along the disk major axis. The peaks of the spectra correspond to the projected Keplerian velocity at the radius. On the southeast side, both \ch{H_2CO} and model spectra (purple and blue lines, respectively) show a symmetric shape. The \ch{H_2CO} spectrum is well reproduced by the best-fit model. In contrast, on the northwest side, where the primary streamer is found in the \ch{HCO^+} emission, the \ch{H_2CO} spectrum is slightly asymmetric and exhibits additional low-velocity components around $V_\mathrm{LSR} - \vsys = 2~\kmps$ that are absent in the best-fit model spectrum. These low velocity components are more prominent in the \ch{HCO^+} emission, and part of the streamer \citep[see Figure 4 in][]{Yen2019a}. The velocity channels corresponding to the additional low-velocity components seen in the \ch{H_2CO} emission were excluded from the fitting, and thus did not affect the line width profile and derived nonthermal velocity dispersions, presented in Figure \ref{fig:dvprof_fitres}.

On the southeast side (or the opposite side to the main streamer), where the \ch{H_2CO} emission traces the Keplerian disk well, the residuals generally remain within $3\sigma$ fitting uncertainties, but also show negative and positive residuals at $\gtrsim3\sigma$ levels. These residuals suggest spatial variations in the line width that are not captured by the current model, which only account for the radial dependence of the local line width. Such variations may arise from an intrinsic azimuthal structure of the velocity dispersion, or from anisotropy in the velocity dispersion, which would induce an azimuthal dependence of the line width through the projection onto POS.

Figure \ref{fig:dvprof_comp} shows radial profiles of the total line width, which are obtained by deprojecting the total line width maps and taking azimuthal average. For the deprojection, we adopted the position and inclination angles from the best-fit model. We only used a half of the disk on the southeast side to avoid contamination of the streamer component. In the upper panel of Figure \ref{fig:dvprof_comp}, the total line widths of the model are scattered at a given radius before azimuthal averaging. This scattering arises from Keplerian shear motion, whose contribution to the line width is not spatially uniform on POS. The total line widths of the data are more scattered than those of the model before azimuthal averaging, which indicates intrinsic spatial variations in the line width in the data in addition to those imposed by the Keplerian shear motions. The residuals after azimuthal averaging are presented in the bottom panel of Figure \ref{fig:dvprof_comp}. After taking the average, statistical uncertainties of the total line widths are $10~\mps$ or even smaller. The residual averaged over radii from 80 to 180 au is nearly zero with the standard deviation of $\rtsim10~\mps$. Hence, we consider that the modeled line width profile well represents the azimuthally averaged line width distribution of the observed data.

For comparison, we conducted another fitting considering only the thermal line broadening. In this fitting, we fixed the parameters regarding the disk geometry and kinematics to the best-fit values in the previous fitting. For gas temperature, the upper limit temperature obtained from the \ce{HCO^+} lines was adopted \citep{Yen2019a}. The local line width was computed with Equations (\ref{eq:dv_def}) and (\ref{eq:vth}) forcing $\vnth = 0$ instead of using Equation (\ref{eq:dv_prof}). We allowed parameters of the gas surface density distribution $(N_\mathrm{gc}, R_\mathrm{c,g},\gamma_\mathrm{g})$ to vary during the fitting to scale the model intensity and examine whether the optical depth compensates for the nonthermal line broadening. Figure \ref{fig:dvprof_secfit_comp} presents comparisons of the radial profile of the total line width between the data and best-fit model without the nonthermal component. The total line widths of the data are clearly larger than those of the best-fit model by $\rtsim75 \pm 15~\mps$. Given the small statistical errors, these offsets suggest that additional nonthermal components is required to reproduce the observations.

\subsection{Effect of Vertical Structure}\label{sec:emission_height}

% ------------ Figure -------------
\begin{figure}
    \centering
    \includegraphics[width=\linewidth]{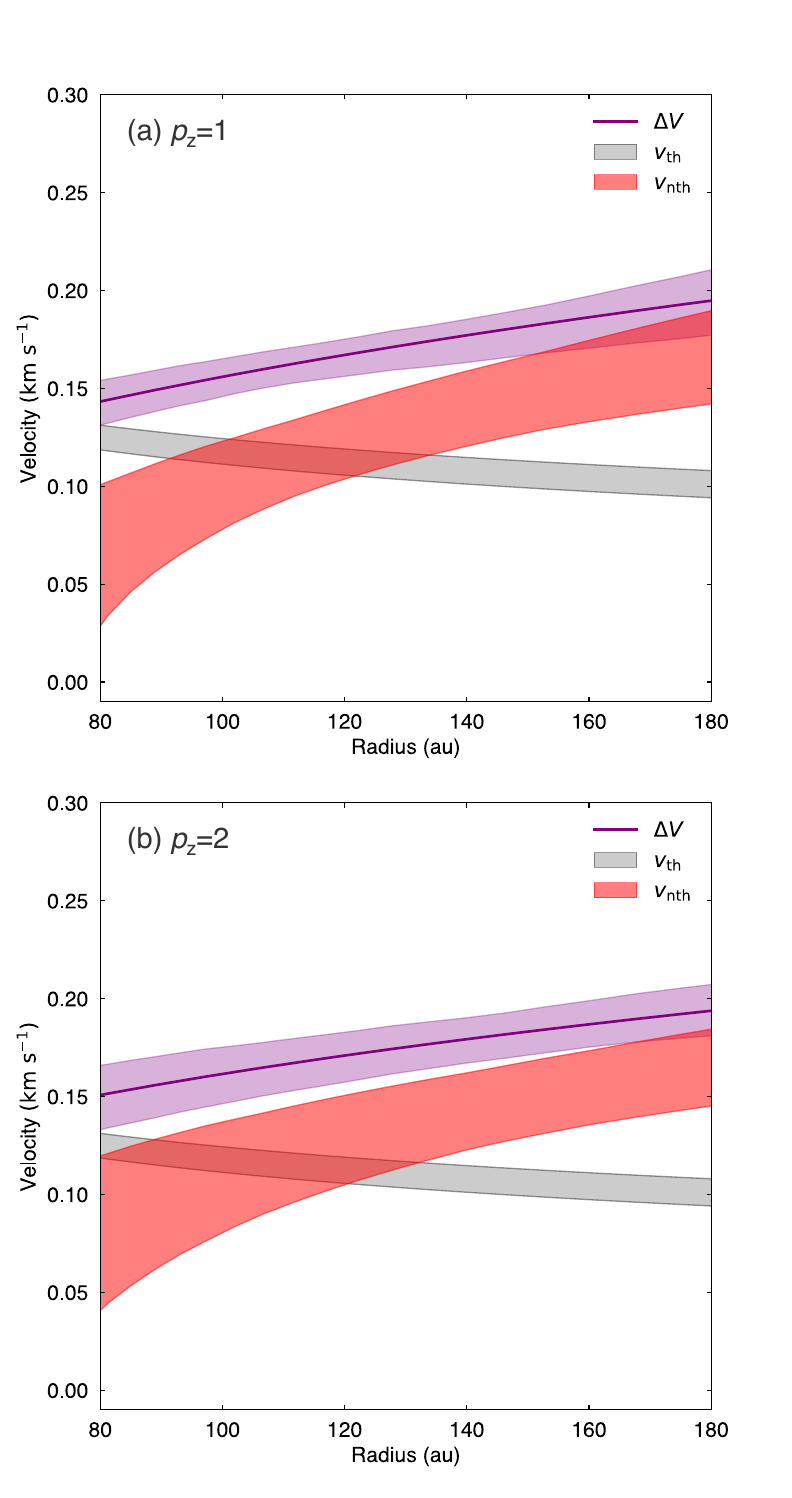}
    \caption{Same as Figure \ref{fig:dvprof_fitres} but for the layered models with $p_z=1$ and $p_z=2$.}
    \label{fig:dvprof_fitres_layered_pz1}
\end{figure}
% ----------------------------------

% ------------ Figure -------------
\begin{figure}
    \centering
    \includegraphics[width=\linewidth]{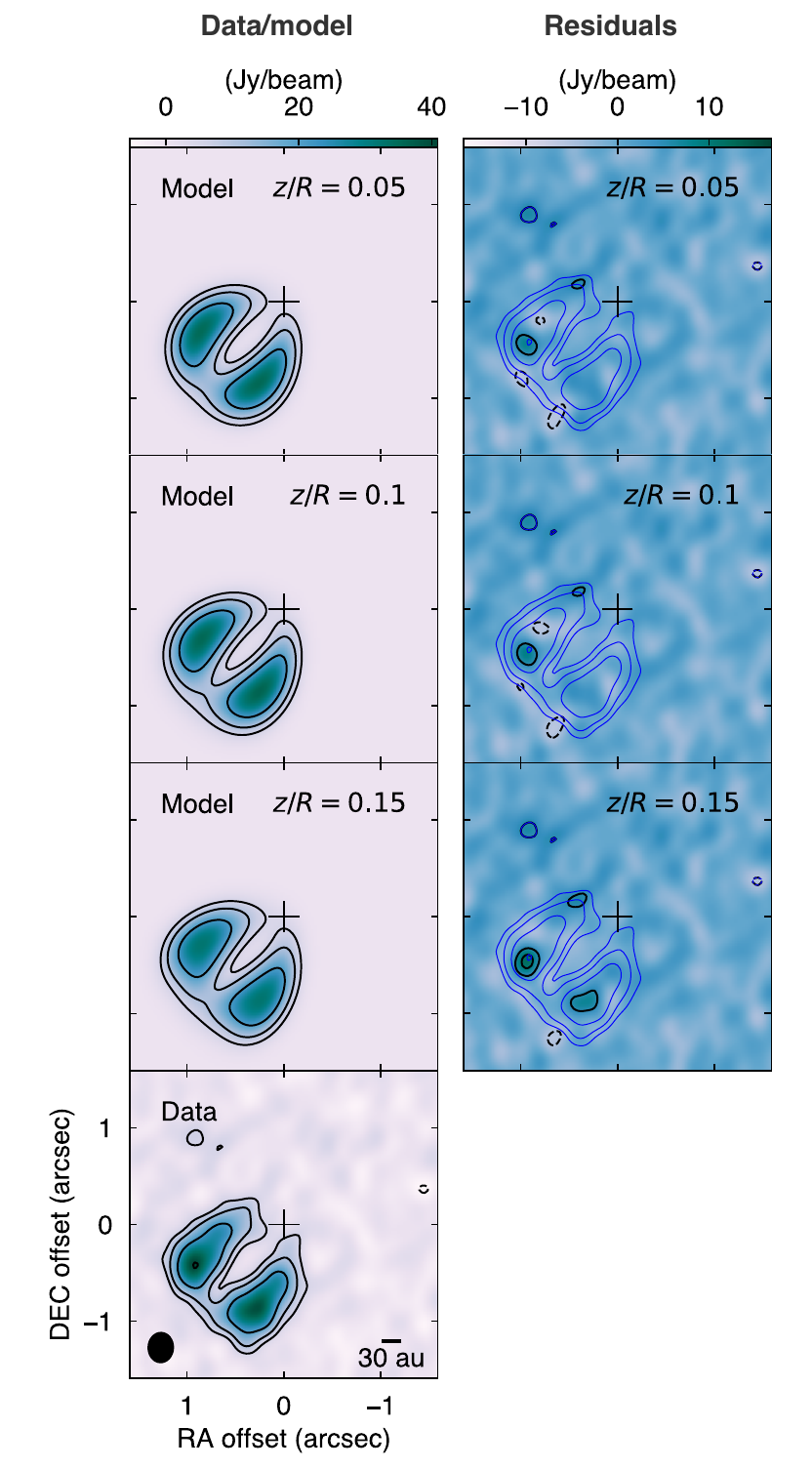}
    \caption{Representative channel maps of the layered models with a fixed flaring index of $p_z=1$ and layer heights of $z/R=0.05$, $0.1$ and $0.15$, compared with the channel map of the \ch{H_2CO} data. The channel maps of the models and data are presented in the left column, while their residuals are shown in the right column. The LSR velocity of the channel is $5.11~\kmps$. Contour levels are $-3\sigma, 3\sigma, 6\sigma, 12\sigma$ and  $24\sigma$, where $\sigma = 1.68~\mjypbm$. Blue contours overlaying the residual maps show the \ch{H_2CO} data.}
    \label{fig:channel_comp_zr}
\end{figure}
% ----------------------------------

The fiducial model presented in the previous section adopts the \ch{H_2CO} density distribution that follows the hydrostatic equilibrium, assuming a constant \ch{H_2CO} abundance throughout the disk. Consequently, the \ch{H_2CO} density peaks at the disk midplane. On the other hand, molecular distributions within protoplanetary disks are often vertically stratified \citep{Law2021b}. The \ch{H_2CO} gas in the HL Tau disk may therefore have a density peak at a finite vertical height of the disk rather than the disk midplane. The vertically elevated emitting layers can compensate for line broadening. Furthermore, the height of the emitting layers provides crucial information about the origin of the nonthermal velocity dispersion. To evaluate the \ch{H_2CO} emission height and its impact on the derived local line width, we fitted the \ch{H_2CO} channel maps with the layered model, whose the \ch{H_2CO} density peaks at a finite vertical height above the disk midplane.

% Model description
In the layered model, the \ch{H_2CO} density structure is modified, while the reminder of the model setup follows the same set of equations as the fiducial model. The \ch{H_2CO} molecules are assumed to reside in vertically elevated layers near the finite height of $z_\mathrm{ml}$:
\begin{align}
    z_\mathrm{ml} & = \pm z_0 \left( \frac{R}{100~\au} \right)^{p_z}, \\
    n_\mathrm{g} &= \frac{N_\mathrm{g}(R)}{\sqrt{2\pi} H_\mathrm{ml}} \exp \left\{- \frac{(z - z_\mathrm{ml})^2}{2H_\mathrm{ml}^2} \right\},
\end{align}
where the thickness of the molecular layers $H_\mathrm{ml}$ is defined as a power-law function
\begin{align}
    H_\mathrm{ml} = H_0 \left(\frac{R}{100~\au} \right)^{p_H}.
\end{align}
The \ch{H_2CO} surface density follows Equation (\ref{eq:gas_surface_density}). As found in these equations, the two \ch{H_2CO} layers are assumed to be symmetric with respect to the disk midplane. The two layers share a common power-law temperature profile (Equation \ref{eq:gas_temp}). We fixed the flaring index $p_z$ to reduce the number of free parameters for better parameter convergence, and examined two cases of $p_z=1$ and $p_z=2$. These prescriptions introduce three additional free parameters ($z_0$, $H_\mathrm{ml}$, $p_H$) compared with the fiducial model. Fits with $p_z = 1$ tend to converge toward small $H_\mathrm{ml}$ values that are not adequately resolved by the model grid. Therefore, when $p_z$ is fixed to unity and $z_0$ is allowed to vary, we impose a lower limit of $H_\mathrm{ml} = 1~\au$ to ensure that the molecular layer retains a finite thickness that is resolved by the model grid. MCMC sampling was run with 32 walkers and 3500 steps, with the first 3000 steps discarded as burn-in.

% Result
The fitting results are summarized in Table \ref{tab:params_layered} in Appendix \ref{app:supp_models}. Both the height of the molecular layer and the local line width are well constrained by the fitting, as seen in the corner plot presented in Figure \ref{fig:corner}. Line broadening arising from the elevated emitting layers and the isotropic velocity dispersion exhibits different spatial dependencies. Because our fitting utilizes full channel maps, this difference would help break the degeneracy between these two effects. The best-fit \ch{H_2CO} layer heights for the models with $p_z=1$ and $p_z=2$ are $z_0 = 4.9\pm0.8~\au$ and $z_0 = 3.7\pm0.6~\au$ at $R = 100~\au$, respectively. Figure \ref{fig:dvprof_fitres_layered_pz1} presents the resulting local line width profile, as well as the radial profiles of the thermal and nonthermal velocity dispersions. The local line width, and consequently the nonthermal velocity dispersion, increase with increasing radius. The same trend is found in the fiducial model, as discussed in Section \ref{sec:fiducial_model}. The average nonthermal velocity dispersion over radii of $80\mbox{--}180~\au$ is $\rtsim0.12~\kmps$, slightly smaller than that obtained with the fiducial model. Nevertheless, it remains significant compared with a statistical uncertainty of $\rtsim30~\mps$, which is obtained through a propagation of the fitting and temperature uncertainties.

For comparison, we also fitted the data with layered models with a fixed flaring index of $p_z=1$ and larger reference heights of $z_0= 10~\au$ and $15~\au$. These models correspond to the emission heights of $z/R=0.1$ and $0.15$, respectively. Although the emission height of $z/R\tsim0.05$ is favored over larger heights based on the likelihood in the MCMC fitting, these additional fits are intended primarily to visualize the differences that arise from different emission heights. The model parameters other than $z_0$ and $p_z$ were reoptimized through MCMC sampling for the models with fixed emission heights of $z/R=0.1$ and $0.15$. MCMC sampling was performed with 30 walkers and 2500 steps with the first 2000 steps discarded as burn-in. The fitting results are summarized in Table \ref{tab:params_layered} in Appendix \ref{app:supp_models}.

Figure \ref{fig:channel_comp_zr} compares representative channel maps of the best-fit layered models with $z/R=0.05$, $0.1$, and $0.15$ with the observed data. The difference in the emission height results in slightly different intensity distributions in each channel map. In Figure \ref{fig:channel_comp_zr}, the model with $z/R=0.15$ reproduces the overall extent of the observed emission, but underpredicts the peak intensity, resulting systemically larger residuals. Increasing the model intensity to match the observed peak intensity in turn broadens the extent of the emission beyond that observed. This visual comparison demonstrates that such large emission heights are disfavored. In contrast, the models with $z/R=0.05$ and $0.1$ show no apparent differences in the channel maps, suggesting that both emission heights are sufficiently small to reproduce the data based on visual inspection alone. Nevertheless, within the current modeling framework, the likelihood is maximized near $z/R\tsim0.05$. We therefore retain $z/R\sim0.05$ as the nominal best-fit value, while emphasizing that higher resolution data are required to spatially resolve the emitting layers and place more robust constraints on the emission height. The model with $z/R=0.1$ yields a local line width of $\Delta V_0 = 0.15 \pm 0.01~\kmps$, which is almost comparable to that obtained for the model with $z/R\tsim0.05$ ($\Delta V_0=0.16\pm0.01~\kmps$). Consequently, a substantial nonthermal velocity dispersion of $\rtsim0.12~\kmps$ is still required to reproduce the observations even if the emission originates from a height of $z/R=0.1$.

The estimated, relatively small emission height of $z/R\sim0.05$ agrees with previous studies of emission heights in Class \II~disks. \cite{Law2021b} have shown that less abundant molecules tend to trace deeper layers close to the disk midplane, based on analyses of CO isotopologues. More recently, \cite{Paneque-Carreno2023a} reported that \ch{H_2CO} emission typically arises from disk heights of $z/R \lesssim 0.1$, based on a sample of several Class \II~sources. These observational results also support our estimate of the \ce{H_2CO} layer height of $z/R\sim0.05$.

\smallskip
In summary, we first fitted the observed \ch{H_2CO} channel maps with the fiducial parametric model of a Keplerian disk, whose the \ch{H_2CO} density distribution follows the hydrostatic equilibrium and peaks at the disk midplane. The model fitting that accounts for Keplerian shear motions suggests an average nonthermal velocity dispersion of $\rtsim0.15~\kmps$ with an increasing radial trend at radii of 80--180 au. The suggested nonthermal velocity is significant compared to the uncertainty of $\rtsim30~\mps$, including statistical errors of $\Delta V$ and uncertainties in the gas temperature. A comparison between the observed and modeled total line width maps reveals spatial variations in the velocity dispersion that are present in the data but not captured by the current model. Nevertheless, the model reproduces the azimuthally averaged line widths of the data within an accuracy of $10~\mps$.

In addition, we fitted the layered model, whose the \ch{H_2CO} density peaks at a finite vertical height above the disk midplane, to examine the \ch{H_2CO} emission height and its impact on the derived nonthermal velocity dispersion. The fit with the layered model suggests that the \ch{H_2CO} emission height is $z/R\tsim0.05$, which is consistent with those estimated in Class \II~disks. The derived average nonthermal velocity dispersion is $\rtsim0.12~\kmps$, which is slightly smaller than that obtained with the fiducial model but still significant compared to the statistical uncertainty.

Taking into account the uncertainty in the vertical density structure of the \ch{H_2CO} gas, we estimate the average nonthermal velocity dispersion over the radii of $\rtsim80\mbox{--}180~\au$ to be $\rtsim0.12\mbox{--}0.15~\kmps$. The derived nonthermal velocity dispersion corresponds to $\rtsim0.5~\cs$ assuming a disk temperature of $20~\kelv$.

\section{Discussion} \label{sec:discussion}

\subsection{Uncertainties} \label{subsec:uncertainties}

% Temperature
The uncertainty of the nonthermal velocity dispersion is estimated to be $\rtsim30~\mps$ based on statistical errors of $\Delta V$ and uncertainties in the gas temperature, which are derived from a temperature range of the outer disk of HL Tau reported in previous studies, as discussed in Section \ref{subsec:measure_vnth}. \cite{Carrasco-Gonzalez2019a} derived a significantly higher dust temperature than the temperature range suggested in other studies. When their dust temperature profile is adopted, an average temperature over radii of $80\mbox{--}180~\au$ is $\rtsim42~\kelv$, yielding a thermal velocity of $\rtsim 0.15~\kmps$. This larger thermal velocity consequently decreases the derived average nonthermal velocity to $0.08\mbox{--}0.13~\kmps$. Nevertheless, the nonthermal component of $\rtsim0.2 \cs$ is still required, where $\cs$ also increases to $\rtsim0.5~\kmps$ as the temperature increases. 

% Systemic uncertainty
In addition to these uncertainties, another possible uncertainty is the systemic error inherited from our modeling approach. To examine the accuracy of the current model fitting approach, we applied the method to mock observational data generated using full radiative transfer models with known turbulent velocities. We fitted the mock data with and without channel masking, which restricts the fit to only a subset of the velocity channels, to assess the robustness of the fitting approach and impact of the channel masking separately. In addition, we also tested how the $uv$ sampling and imaging with CLEAN affect the fitting result with an ALMA simulation. Details of these tests are presented in Appendix \ref{app:radmcmodel}, and the results of measurements of the nonthermal velocity dispersions using the mock data are provided in Figure \ref{fig:dvprof_RTmodel} in the appendix. In these experiments, which adopt the same density structure of the disk, the nonthermal velocity is recovered to within an accuracy of $\rtsim20~\mps$ when $\vnth = 0.3 \mbox{--}0.5~\cs$, regardless of the $uv$ sampling and channel masking. Hence, the measured nonthermal velocity dispersion of $0.5~\cs$ appears robust against these uncertainties.

The tests presented in Appendix~\ref{app:radmcmodel} also demonstrate that a nonthermal velocity dispersion of $0.5~\cs$ can be measured at the spectral resolution of the present data through forward modeling of the entire spectral cube. Nevertheless, we note that the inferred nonthermal velocity dispersion is slightly smaller than the spectral resolution of $0.187~\kmps$ and is close to the practical limit of what can be robustly constrained from the present data.

\subsection{Origin of the large nonthermal velocity dispersion} \label{subsec:origin}

Our analysis suggests a large nonthermal velocity dispersion of $\vnth \tsim 0.5~\cs$ near the midplane of the outer disk of HL Tau. In this section, we discuss the physical origin of the large nonthermal velocity dispersion.

\subsubsection{Turbulence}

Turbulence contributes to the line broadening, as we attempted to measure it through this work. Assuming that the measured nonthermal velocity dispersion is purely due to turbulence, we obtain a turbulent Mach number $\mathcal{M} \tsim0.4$ in the outer disk of HL Tau. This can be further translated into the dimensionless parameter $\alpha\tsim0.16$ via $\alpha \tsim \mathcal{M}^2$. Several mechanisms have been proposed as the origin of turbulence in protoplanetary disks \citep{Lyra2019a, Lesur2023a} such as the magnetorotaional instability \citep[MRI; ][]{Balbus1991a}, vertical share instability \citep[VSI; ][]{Nelson2013a, Lin2015a, Fukuhara2021a} and the gravitational instability \citep[GI; ][]{Kratter2016a}. Infall is also a possible driving source of turbulence, as the HL Tau disk is associated with an infalling envelope and streamers \citep{Lesur2015a, Hennebelle2017a, Kuznetsova2022a}.

% MRI
Local numerical simulations of MRI show that the MRI-driven turbulence can be as strong as $\mathcal{M}\tsim 0.4$ or $\alpha \tsim 0.1$ at high disk altitudes ($z/R > 0.2$), but is typically much smaller than $0.1~\cs$ near the disk midplane ($z/R < 0.1$) \citep{Okuzumi2011a, Simon2018a}. The MRI can drive a strong turbulence of $\mathcal{M}\tsim0.4$ even near the disk midplane when the magnetic field is strongly coupled with gas in a condition nearly ideal MHD \citep{Bai2011a, Okuzumi2011a, Simon2018a}. However, such a condition is usually not expected in the midplane of protoplanetary disks at a large radius of $\rtsim 100~\au$ because of the low ionization degree of the disk gas.
%the magnetic field is very strong as the plasma $\beta < 10$ and 
Similarly, numerical simulations suggest that VSI drives turbulence of $\rtsim0.1~\kmps$, or $\mathcal{M}\tsim0.3$ at most \citep{Flock2020a, Lesur2025a}, which is smaller than the value suggested by this work. Hence, it would be difficult to explain the observed large velocity dispersion with MRI or VSI turbulence.

% GI
GI possibly explains the large velocity dispersion. GI drives turbulence as strong as $\vturb \tsim 0.2\mbox{--}0.6~\cs$ near the disk midplane \citep{Shi2014a, Riols2020b}, which matches the measured nonthermal velocity. Indeed, previous observational work suggests that the HL Tau disk is subject to GI.
The criterion whether GI occurs or not is often assessed with the Toomre $Q$ parameter \citep{Toomre1964a}:
\begin{align}
    Q = \frac{\cs \kappa}{\pi G \Sigma},
\end{align}
where $\Sigma$ is the gas surface density, and $\kappa$ is the epicyclic frequency that is equal to the angular velocity $\Omega$ in Keplerian disks. 
Based on observations in the dust continuum emission, previous studies estimated the Toomre $Q$ parameter to be larger than 2 at inner radii but close to 2 near the edge of the dusty disk, inferring that the outer disk may be GI unstable \citep{Akiyama2016a, Liu2017a}. A more robust measurement of the disk mass using an optically thin \ce{^{13}C^{17}O} emission suggests $Q < 1.7$ at radii of $50\mbox{--}100~\au$ and that the outer disk of HL Tau is gravitationally unstable \citep{Booth2020a}.

The Toomre $Q$ parameter generally decreases with radius, given the radial dependence of typical temperature and surface density profiles of protoplanetary disks, and Keplerian angular velocity. Consequently, it is naturally expected that a disk becomes GI unstable toward outer radii. This tendency also aligns with the fact that turbulence is as weak as $\alpha \leq 10^{-4}$ in the inner disk of HL Tau \citep{Pinte2016a, Yang2025a}, i.e., implying that the disk is gravitationally stable in the inner (low $\alpha$) region and unstable in the outer (high $\alpha$) region.

The possibility that the outer disk is gravitationally unstable is further supported by the expectation that the disk is in an increasing phase of disk mass.
The mass infall rate from the envelope to the disk in HL Tau has been estimated to be $2.2 \times 10^{-6}~\msunpyr$ based on an infalling arc structure on $1000~\au$ scale, detected in the \ch{C^{18}O} emission \citep{Yen2017b, Yen2019b}. 
The mass accretion rate onto the protostar is estimated to be $1.6\times10^{-7}~\msunpyr$ based on a spectroscopy \citep{White2004a}, which is smaller than the mass infall rate by an order of magnitude. Hence, the disk should grow on a timescale of $0.1~\mathrm{Myr}$ to reach a GI unstable regime, $\Mdisk/M_\ast\tsim0.1$.

The mass accretion rate need not be spatially uniform across the disk, particularly when substructures are formed by pressure maxima that traps material. If the outer disk is gravitationally unstable, it is expected to sustain a higher mass accretion rate than the inner regions. In the viscous disk frame work, the mass accretion rate can be expressed as
\begin{align}
    \macc = 3 \pi \Sigma \alpha \frac{\cs^2}{\Omega}.
    \label{eq:Macc_viscous}
\end{align}
The surface density at the characteristic radius $\Sigma_\mathrm{c}$ can be written with the disk mass as
\begin{align}
    \Sigma_\mathrm{c} = \frac{(2 - \gamma) \Mdisk}{2\pi R_\mathrm{c}^2}.
    \label{eq:sig_c}
\end{align}
The mass accretion rate at the characteristic radius can be estimated using these equations. 
We adopt $R_\mathrm{c} = 100~\au$, $\gamma = -0.7$ and $T=25~\kelv$ from our best-fit model and temperature estimates by previous work.
Assuming $\alpha \tsim0.1$ and $\Mdisk = 0.2~\Msun$ \citep{Booth2020a}, we obtain $\macc \tsim 3.3\times10^{-6}~\msunpyr$. This value is broadly consistent with mass accretion rates driven by angular momentum transport via GI-induced spiral structures in numerical simulations at evolutionary stages comparable to the Class~I phase \citep{Vorobyov2007a}, and also with the observationally estimated value of the mass infall rate from the envelope.

Gravitational instability is a self-regulating process: the development of spiral arms and the subsequent generation of (self-gravitating) turbulence drive mass accretion, which reduces the disk surface density and thereby stabilizes the disk. As a result, the instability (and turbulence) is suppressed on a timescale of the local orbital period, and renewed mass supply is required for the instability to grow again and maintain the turbulence.
For HL Tau, the orbital timescale at $R=100~\au$ is $t_\mathrm{rot} = 2 \pi \sqrt{R^3 /(G M_\ast)} \tsim 650~\yr$ assuming the protostellar mass of $2.34~\Msun$. Combined with the mass accretion rate estimated above, this orbital timescale implies a mass of $\rtsim2~M_\mathrm{Jup}$ is removed from the outer disk before the spirals decay and the outer disk temporarily returns a GI-stable state. A comparable mass infall rate from the envelope onto the disk, estimated to be $2.2\times 10^{-6}~\msunpyr$, suggests that the depleted mass can be replenished on a timescale of $\rtsim1000~\yr$, allowing the disk to become a GI-unstable phase again. This recurrence timescale is much shorter than the typical lifetime of Class I objects \citep{Evans2009a}, which would make such states naturally observable.

Infall from the envelope onto the disk and its dynamical perturbation may be another possible origin of strong turbulence. Hydrodynamical simulations show that accretion shock between the disk and infalling gas generates spiral density waves, leading to non-negligible Reynolds stress $\alpha$ throughout the disk \citep{Lesur2015a, Hennebelle2017a,
Kuznetsova2022a}. The Reynolds stress induced by such spiral waves can be as strong as $\alpha~\tsim0.1\mbox{--}1$ near the edge of the disk and propagate to smaller radii decreasing down to $\alpha\tsim10^{-3}\mbox{--}10^{-4}$ \citep{Lesur2015a, Hennebelle2017a}. These high $\alpha$ values in the outer disk and a increasing radial trend are consistent with the results of our analysis. These theoretical studies assume disk masses and mass infall rates on the orders of $10^{-2}~\Msun$ and $10^{-7}~\msunpyr$, respectively, both of which are an order of magnitude smaller than those estimated for HL Tau. Nevertheless, when considering the ratio of the mass infall rate to the disk mass ($\dot{M_\mathrm{inf}} / \Mdisk$), the parameters explored in the numerical simulations are comparable to the values for HL Tau. Hence, the infalling gas could induce density perturbations of similar magnitude to those seen in the simulations and potentially drive turbulence of $\alpha\tsim0.1$ in the outer disk of HL Tau.

A caveat of both GI- and infall-driven turbulence scenarios is that no clear spiral structures have been identified in the HL Tau disk, even though they are expected to drive the strong velocity perturbations and angular momentum transport. One possible explanation is that dust grains have already drifted inward, and spiral structures at outer radii may not be detectable in the dust continuum. Velocity perturbations associated with GI-induced spirals are typically confined to spatial scales of $\lesssim10~\au$ at a radius of $\rtsim100~\au$ \citep{Hall2020b,Sai2025a, Barraza-Alfaro2025a}. The current angular resolution of the \ch{H_2CO} data is therefore insufficient to resolve such features, even if they are present. Future observations at high angular resolutions in molecular lines would shed light on the presence or absence of spiral structures in the outer disk.

\subsubsection{Non-turbulence origin}
% if turbulence or not
One may wonder whether the disk wind contributes to the line broadening rather than turbulence, as the bipolar outflow has been observed in HL Tau \citep{Cabrit1996a, Takami2007a}. The velocity structures of the \ch{H_2CO} emission mostly follow Keplerian rotation of the disk and do not show any features of outflowing gas motions, as seen in Section \ref{subsec:obs_result} and Section \ref{sec:fitres}. The base of the disk wind could contaminate and contribute to the total line width. However, the base of the disk wind is typically located at the FUV ionized surface at $z/H \gtrsim 5$ \citep{Bai2013a, Gressel2015a, Lesur2023a},
which is much higher than the layer traced by the \ch{H_2CO} emission. Furthermore, it is expected that the gas density at the wind base is too low to contribute the line emission. Hence, it is unlikely that the disk wind is responsible for the observed nonthermal velocity dispersion. 

Radial gas flows within the disk plane are naturally expected in the accretion disk, which could contribute to the line broadening but is typically negligible \citep{Pinte2022b}. In the frame work of the viscous disk model \citep{Shakura1973a}, the radial accretion velocity within the disk is expressed as
\begin{align}
    v_R = \frac{3}{2} \alpha \left(\frac{H}{R}\right)^2\vkep,
    \label{eq:vr_viscous}
\end{align}
where $\vkep$ is the Keplerian velocity. For HL Tau, the Keplerian velocity is $\rtsim4.6~\kmps$ at a radius of $100~\au$. Even assuming a highly turbulent case of $\alpha\tsim0.1$, the accretion velocity at a radius of $100~\au$ is estimated to be $v_R\tsim7~\mps$ with the typical aspect ratio of the disk of $H/R\tsim0.1$, which is much smaller than the suggested nonthermal velocity dispersion.

As the HL Tau disk possesses multiple rings in the dusty disk and possibly in the gas disk \citep{ALMAPartnership2015b, Yen2019a}, non-Keplerian motion may be induced by the change in the pressure gradient around the gaps and contribute to the line broadening \citep{Teague2018a}. The non-Keplerian motions caused by the pressure gradient term are typically a few percent of the projected Keplerian speed \citep{Teague2018a, Pinte2022b}. In the case of HL Tau, the velocity of $2\mbox{--}3\%$ of the projected Keplerian at outer gap locations of $R=80~\au$ and $100~\au$ \citep{Stephens2023a} is $\rtsim0.06\mbox{--}0.1~\kmps$, which is insufficient to explain the observed nonthermal line broadening alone. Furthermore, in the HL Tau disk, gaps possibly causing these velocity deviations have been confirmed only at radii smaller than $100~\au$, while the \ch{H_2CO} emission extends up to a radius of $\rtsim200~\au$. Hence, our measurements that are more sensitive to the outer disk would not be significantly affected by these possible non-Keplerian motions associated with the known gaps.

In addition to the non-Keplerian motions associated with gas gaps, several Class \II~disks exhibit localized kinematic perturbations on spatial scales of a few tens au \citep{Casassus2019a, Teague2021a,Teague2022a}. If such motions are present in the outer disk of HL Tau, they would be smeared out by the current spatial resolution of $\rtsim40~\au$ and could contribute to the observed line broadening. Molecular line observations at higher angular resolutions are required to further investigate this possibility.

\subsection{Comparison with turbulence estimates within the dusty disk}\label{subsec:comp_with_dust}

\begin{deluxetable*}{lcccccc}
    %\digitalasset
    %\tablewidth{0.8\columnwidth}
\tablecaption{Comparisons with measurements of turbulence via dust scale heights \label{tab:summary_turb}}
\tablehead{
\colhead{Measured quantity} & 
\colhead{Measured value} & \colhead{Data} &
\colhead{Radius} & \colhead{Height} & \colhead{Inferred $\alpha^1$} & \colhead{References}
 \\
\colhead{} & \colhead{} & \colhead{} & \colhead{(au)} & \colhead{($R$)} & \colhead{} & \colhead{} }
\startdata
Dust scale height ($H_\mathrm{d}/R$) & $\Hd / R \sim0.01$ & $0.87\mbox{--}2.9$ mm cont. & $\lesssim100$ & $\sim0.01$ &  $\rtsim10^{-4}$ & (1) \\
Dust scale height ($H_\mathrm{d}/R$)& $\Hd / R \gtrsim 0.1$ & 0.87 mm polarized cont. & $\rtsim20$ & $\gtrsim0.1$ & $\rtsim10^{-2.5}$ & (2) \\
Dust scale height ($H_\mathrm{d}/R$)& $\Hd / R < 0.05$ & 0.87 mm polarized cont. &  $40\mbox{--}120$ & $<0.05$ & $\rtsim10^{-5}$ & (2) \\
Line width ($\vturb$) & $\vturb\tsim0.5~\cs$ & \ch{H_2CO} ($3_{1,2}\mbox{--}2_{1,1}$) & $80\mbox{--}180$ & $\rtsim0.05$ & $\sim0.16$ & (3) \\
\enddata
\tablecomments{References: (1) \cite{Pinte2016a}; (2) \cite{Yang2025a}; (3) This work. $^\mathrm{1}$ $\alpha \tsim \mathcal{M}^2$ is assumed when it is inferred from the turbulent velocity.}
\end{deluxetable*}

Previous studies of HL Tau placed constraints on the turbulence strength with ALMA millimeter continuum observations. \cite{Pinte2016a} has modeled the ALMA dust continuum emission at 0.87--2.9 mm wavelengths to measure the dust vertical scale height, finding that the dust scale height is as small as $\Hd/R\rtsim0.01$ at radii of $\lesssim 100~\au$. A following work done by \cite{Yang2025a} has measured the dust scale height of each dusty ring using the Stokes $Q$ of the polarized continuum emission. They reported a radial variation in the dust scale height with a higher value at an inner radius of $20~\au$ and smaller heights of $\Hd/R < 0.05$ in outer rings, the latter of which agrees with the measurement by \cite{Pinte2016a}. These small dust scale heights at radii of $40\mbox{--}120~\au$ of HL Tau and dust diffusion models yield turbulence strengths of $\alpha \rtsim 10^{-4}$ or even smaller with (sub)millimeter dust grains. Table \ref{tab:summary_turb} summarizes these previous measurements based on the dust continuum observations in comparison with the results of this work.

The strength of turbulence measured in the present work is larger than those found in previous work by more than three orders of magnitude. A possible explanation for this discrepancy is a vertical gradient in the turbulence strength. VSI and MRI can drive turbulence with a vertical gradient \citep{Simon2018a, Fukuhara2023a}. However, in VSI and MRI models where the turbulence remains weak near the disk midplane, turbulent velocities of $\mathcal{M}\gtrsim0.3~\cs$ are generally reached only at high disk altitudes of $z/R\gtrsim 0.2$ \citep{Simon2018a, Fukuhara2023a}. Because our analysis suggests that the \ch{H_2CO} emission traces gas near the disk midplane ($z/R\tsim0.05$), the vertical gradients of turbulence predicted by these models are not sufficiently steep to explain the discrepancy between the previous and current measurements. Another possibility is that previous studies and the current work trace different disk radii. Our analysis is more sensitive to outer regions of $R\tsim100\mbox{--}200~\au$ than the inner disk of $R\lesssim100~\au$, where the dust continuum emission of previous studies mainly traces. In fact, the best-fit nonthermal velocity dispersion increases with radius under the assumption of a smooth power-law function.
Anisotropic turbulence could also explain the differences between the previous and present measurements. The two methods using the line width and dust scale height can trace orthogonal components of turbulent velocities. Our method using the velocity dispersion of the molecular line traces both vertical and horizontal velocity dispersions, $\delta v_z$ and $\delta v_{R\phi}$, given the moderate inclination angle of the disk. In contrast, the dust scale heights reflect vertical dust diffusion caused by $\delta v_z$. Hence, the difference in the strength of turbulence between the previous and current work could be due to in-plane turbulent velocity larger than vertical component, $\delta v_{R\phi} \gg \delta v_z$.

Turbulence within protoplanetary disks can be anisotropic \citep{Riols2020b, Baehr2021a, Fukuhara2023a}. In particular, according to hydrodynamical simulations, GI-induced turbulence is anisotropic with stronger in-plane motions ($\delta v_z \ll \delta v_{R\phi}$) near the disk midplane \citep{Riols2020b, Baehr2021a}. Furthermore, \cite{Baehr2021a} reproduced the observed dust scale height of the HL Tau disk with hydrodynamical models of a GI disk, owing the anisotropic nature of the turbulence and self-gravity of the dust grains. \cite{Jiang2024a} estimated a higher turbulence of $\alpha \tsim 3 \times 10^{-3}$ near an outer ring at $R\tsim85~\au$, based on the size distribution of dust grains in the HL Tau disk and an assumption that the maximum grain size is determined by the fragmentation limit.

The radial coverage of our measurement partially overlaps with the radial range of the previous measurements of turbulence based on the dust scale height. Anisotropic nature of the turbulence could explain this discrepancy. However, we note that the partial overlap at radii of $\rtsim80\mbox{--}120~\au$ could also be due to the limited spatial resolution of the current data and the restricted flexibility of the parametric model. The spatial resolution of the current data is $\rtsim40~\au$, and the radial dependence of the local line width is implemented as a smooth power-law function. Because of these restrictions, our analysis is not sensitive to sharp radial variations in the turbulent velocity on scales of a few tens of au. If a sharp transition between a quiescent inner disk and a turbulent outer disk exists, the present data and model would likely smear out such a feature. Therefore, the inferred turbulent velocity should be considered as an averaged value representative of the outer disk rather than a well-constraint radial profile. Higher angular resolution data are required to investigate the radial distribution of the turbulent velocity in more detail.

\subsection{Comparison to Measurements in Class \II~Disks} \label{subsec:comp}

Direct measurements of local, nonthermal motions using similar modeling approaches have been conducted toward six Class \II~disks so far, among which four disks show low turbulence with upper limits of $\lesssim 0.1\cs$ and two disks (DM Tau and IM Lup) exhibit strong turbulence \citep{Flaherty2020a, Flaherty2024a, Paneque-Carreno2024a, Hardiman2026a}. For the DM Tau disk, \cite{Flaherty2020a} measured turbulence of $0.25\mbox{--}0.33~\cs$ based on the CO $J=2\mbox{--}1$ emission. More recently, \cite{Hardiman2026a} analyzed the CO $J=3\mbox{--}2$ emission toward the same source with a more comprehensive radiative transfer model, reporting a slightly higher value of $0.4~\cs$. For IM Lup, turbulent velocities of $0.18\mbox{--}0.3~\cs$ and $0.5\mbox{--}0.7~\cs$ have been reported with CO isotopologues \citep{Flaherty2020a} and the CN emission tracing a height of $\rtsim2 H$ \citep{Paneque-Carreno2024a}, respectively. The turbulent velocity  measured by this work $\vnth=0.5~\cs$ is comparable to the latter measurement for IM Lup, although the traced emission heights are different. While the origin of these turbulent motions detected in the Class \II~disks are still under debate, the IM Lup disk has been suggested to be under GI because of its large disk mass and two symmetric spiral arms \citep{Andrews2018a, Yoshida2025a, Ueda2024a}. Hence, the similar turbulent strengths inferred for the two sources may be consistent with GI- or infall-driven scenarios for HL~Tau, in which turbulence is generated by spiral density waves. Currently, HL Tau is the only source associated with an infalling envelope, toward which turbulent velocity has been estimated. Future studies of more numbers of young sources would hint at an evolutionary trend of turbulence in protoplanetary disks and also roles of infall in driving turbulence.

%\begin{comment}
\subsection{Implication for Planet Formation} \label{subsec:implication}
Given the multiple, settled dusty rings of the inner disk of HL Tau, the strong nonthermal motions in the outer disk suggests a large variation in the strength of turbulence across the disk radii, and/or anisotropic nature of turbulence. The rings and gaps of the HL Tau disk could be explained by pressure bumps produced by embedded planets \citep{Dong2015b}, although their origins are still under debate. High-density rings trapped by the pressure bump can be suited sites for planet formation \citep{Carrera2021a}. Strong turbulence in the outer disk suggests a high mass accretion rate from the outer disk to the dusty rings possibly pressure-trapped. With the mass accretion rate in the outer disk estimated with Equation (\ref{eq:Macc_viscous}), about two Jupiter masses would be fed to an outermost ring within the local orbital timescale of $\rtsim650~\yr$. The combination of pressure bumps and strong turbulence in the outer disk could foster the planet formation process within a pressure-trapped ring by efficiently accumulating mass, and the HL Tau disk could be a unique case to study such a planet formation process.

\section{Conclusions} \label{sec:conclusion}

We presented the first direct measurement of the nonthermal velocity dispersion in an embedded disk associated with an infalling envelope and streamers, using ALMA archival data of the \ch{H_2CO} ($3_{1,2}\mbox{--}2_{1,1}$) line of HL Tau. With the parametric model fitting approach that accounts for Keplerian shear motions, we detected a nonnegligible nonthermal velocity dispersion of $\rtsim0.12\mbox{--}0.15~\kmps$ on average over radii of $80\mbox{--}180~\au$ in the HL Tau disk with an increasing radial trend. These nonthermal velocity dispersions correspond to $\rtsim0.5~\cs$ or $\mathcal{M}\tsim0.4$, assuming a typical disk temperature of $T=20~\kelv$. In addition, fitting with the layered model, whose the \ch{H_2CO} density peaks at a finite vertical height above the disk midplane, suggests that the height of the \ch{H_2CO} emitting layers is $z/R\rtsim0.05$. Turbulence driven by GI or infall most naturally explains the large nonthermal velocity dispersion near the disk midplane among several possibilities including nonturbulent mechanisms. Observations in molecular lines at high angular resolutions would enable further investigation of potential spiral structures in the outer disk, which are supposed to be present if GI or infall induces strong turbulence via spiral density waves. Regardless of its origin, the strong turbulence in the outer disk, in contrast to weak turbulence within the well settled dusty rings, could play a significant role in the planet formation process by efficiently accumulating mass from the outer disk to the dusty ring. 

\begin{acknowledgments}
This paper makes use of the following ALMA data: ADS/JAO.ALMA\#2018.1.01037.S, ADS/JAO.ALMA\#2016.1.00366.S, and ADS/JAO.ALMA\#2011.0.00015.SV. ALMA is a partnership of ESO (representing its member states), NSF (USA), and NINS (Japan), together with NRC (Canada), NSTC and ASIAA (Taiwan), and KASI (Republic of Korea), in cooperation with the Republic of Chile. The Joint ALMA Observatory is operated by ESO, AUI/NRAO, and NAOJ. We thank all ALMA staff for supporting this work. We thank the anonymous referee for providing an insightful review that helps to improve the paper. JS is supported by JSPS KAKENHI grant Nos.~JP26K17198 and JP26K00741. JS acknowledges support from the KU-DREAM program of Kagoshima University. JS thanks Pin-Gao Gu for insightful feedback on the interpretations of the results from a theoretical point of view. ST is supported by JSPS KAKENHI grant Nos.~JP21H00048, JP21H04495, and JP24K00674, and by NAOJ ALMA Scientific Research grant No.~2022-20A. HWY acknowledges support from the National Science and Technology
Council (NSTC) in Taiwan through the grant NSTC 114-2112-M-001-017-
and from the Academia Sinica Career Development Award
(AS-CDA-111-M03). YT is supported by JST FOREST grant number JPMJFR2234. Data analysis was in part carried out on the Multi-wavelength Data Analysis System operated by the Astronomy Data Center (ADC), NAOJ.

\end{acknowledgments}

\begin{contribution}

JS conducted all the analyses and was responsible for writing and submitting the manuscript. ST provided feedback on interpretations of the results and writing. HWY came up with the initial research concept, and provided the self-calibrated \ch{HCO^+} visibility data and  valuable feedback on analyses. YT and YF helped interpret the results from a theoretical perspective.

\end{contribution}

\facilities{ALMA}

\software{CASA \citep{McMullin2007a}, emcee \citep{Foreman-Mackey2013a}, Numpy \citep{Harris2020a}, Scipy \citep{Virtanen2020a}, Matplotlib \citep{Hunter2007a}, Astropy \citep{TheAstropyCollaboration2013a, TheAstropyCollaboration2018a}. 
}

\appendix

\section{Robustness of the parametric modeling approach}\label{app:radmcmodel}

% ------------ Figure -------------
\begin{figure*}
    \centering
    \includegraphics[width=\linewidth]{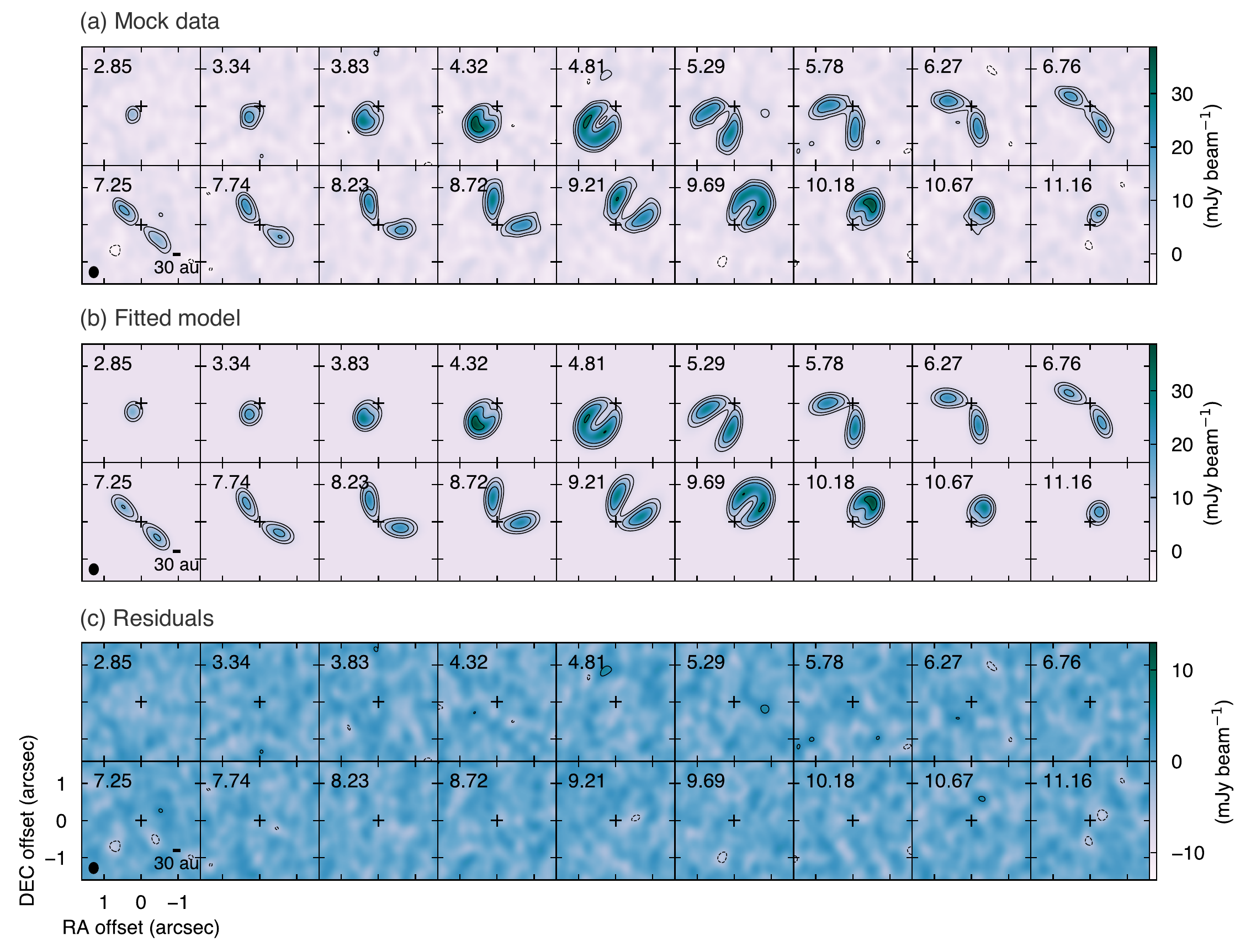}
    \caption{Same as Figure \ref{fig:channel_fitres} but for the mock data and best-fit model in the case of $\vturb = 0.1\cs$.}
    \label{fig:channel_RTmodel}
\end{figure*}
% ----------------------------------

% ------------ Figure -------------
\begin{figure*}
    \centering
    \includegraphics[width=\linewidth]{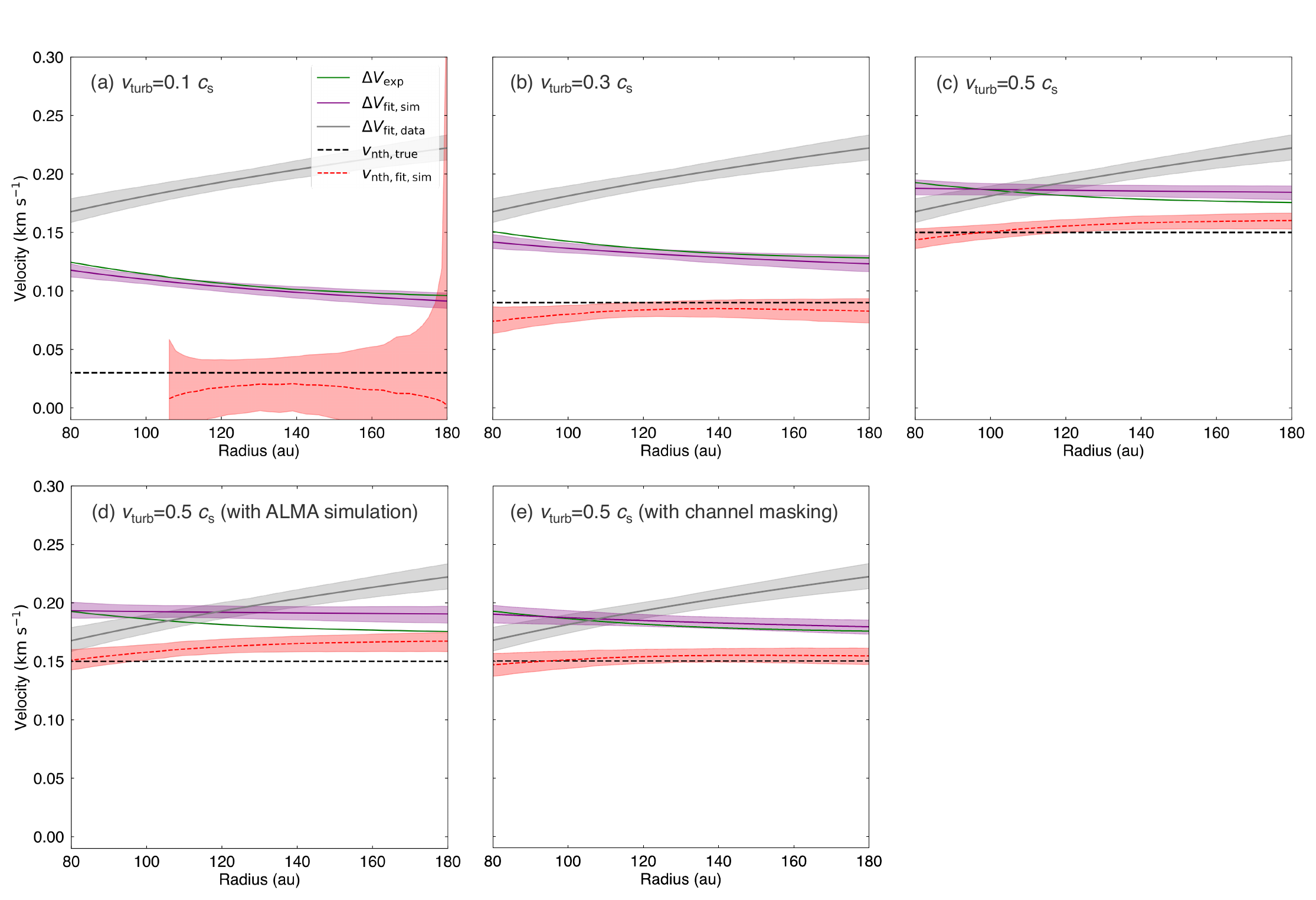}
    \caption{Radial profiles of the local line width expected in the mock data ($\Delta V_\mathrm{exp}$), which is computed with the disk midplane temperature and the input nonthermal velocity, the best-fit local line width for the mock data ($\Delta V_\mathrm{fit,sim}$), the nonthermal component ($v_\mathrm{nth,true}$) imposed in the simulated model, and the nonthermal component derived with the fitted local line width and the midplane temperature of the simulated model ($v_\mathrm{nth,fit, sim}$). For comparison, the local line width derived by the fitting with the actual \ch{H_2CO} data ($\Delta V_\mathrm{fit,data}$) is also plotted. Shaded areas show uncertainties corresponding to $1\sigma$ fitting errors. The isothermal sound speed is assumed to be $\cs = 0.3~\kmps$.}
\label{fig:dvprof_RTmodel}
\end{figure*}
%local line widths and nonthermal velocity dispersions that are assumed the full radiative transfer models and derived through the parametric model fitting with the mock data. Each curves show 
% ----------------------------------

To examine the robustness of the parametric modeling approach, we applied the parametric fitting method to mock line data, which we generated through the full radiative transfer calculation using the public radiative transfer code, RADMC-3D\footnote{\url{https://www.ita.uni-heidelberg.de/∼dullemond/software/radmc-3d/}}. The disk model that was used to generate mock data followed the density distribution described in Section \ref{sec:modeling} but shared the same $R_\mathrm{c}$ and $\gamma$ for both dust and gas. We adopted $R_\mathrm{c} = 84.2~\au$ and $\gamma=-0.2$ from \cite{Kwon2015a} with a correction for the updated distance. The surface density at the critical radius ($\Sigma_\mathrm{c}$) is set such that the total disk mass was $0.15~\Msun$. The disk vertical structure is assumed to follow a power-law function $H \propto r^{1.25}$ with $H_\mathrm{1au} = 0.07~\au$. The gas-to-dust mass ratio and abundance of the \ce{H_2CO} molecule with respect to \ce{H_2} gas were assumed to be 100 and $1\times10^{-10}$ \citep{Oberg2017a}, respectively. The DSHARP opacity table was adopted for the dust opacity \citep{Birnstiel2018a}. We produced three models with different turbulent velocities of $0.03~\kmps$, $0.09~\kmps$ and $0.15~\kmps$, which are assumed to be uniform over the entire disk extent. Considering the typical disk temperature of $20\mbox{--}30~\kelv$ around a radius of $100~\au$, the assumed turbulent velocity in each model approximately corresponds to $0.1~\cs$, $0.3~\cs$ and $0.5~\cs$, where $\cs=0.3~\kmps$.

The two dimensional thermal structure of the disk was computed with RADMC-3D considering irradiation from the central star, which is approximated to be a point source with a blackbody temperature of 5300 K assuming the stellar luminosity of $11~L_\odot$ \citep{Liu2017a}, $R_\ast = 4 R_\odot$ \citep{Stahler1980a} and the Stefan--Boltzmann law. Then, the radiative transfer was calculated with an inclination angle of $46^\circ$ and a position angle of $138^\circ$ \citep{ALMAPartnership2015b}. The channel spacing was set to $0.163~\kmps$. After the radiative transfer calculations, a Gaussian beam with the same size as that of the data was convolved. We also added rms noise of $1.3~\mjypbm$ to match S/N of the model images with that of the observed data. In addition to these beam convolved maps, we also performed ALMA simulation with CASA, adopting an antenna configuration same as that of the actual observations, to examine the impact of the $uv$-sampling and imaging with CLEAN for the case of $\vturb = 0.5~\cs$.

We performed the fitting using the mock data in the same manner as described in Section \ref{subsec:fitting}. First, we fitted the mock continuum data to determine the dust optical depth, and then fitted the mock \ch{H_2CO} line data. To obtain the nonthermal velocity dispersion from the local line width, a temperature profile must be provided. We adopted the disk midplane temperature from the model to subtract the thermal component. Uncertainties associated with errors in temperature structures are discussed separately in Section \ref{subsec:uncertainties}. In this Appendix, we focus on investigating possible systemic uncertainties induced by the fitting process. To assess the baseline performance of the method, we performed the fitting using the full channel maps without applying channel masking. To investigate the impact of channel masking, we also carried out a separate fit for the case of $\vturb = 0.5~\cs$ without ALMA simulation, excluding the same velocity channels that were omitted in the fitting of the HL Tau data.

Figure \ref{fig:channel_RTmodel} presents the channel maps of the simulated data with $\vturb = 0.1\cs$, the best-fit model and residuals after subtracting the best-fit model from the simulated data. The close agreement between the simulated data and the best-fit model found in these channel maps demonstrates that the channel-based three-layer approximation is sufficient to reproduce the results of the full radiative transfer. Figure \ref{fig:dvprof_RTmodel} presents the best-fit radial profiles of the local line width (purple curves). Purple shaded areas indicate the $1\sigma$ fitting uncertainty of $\Delta V$, derived from the posterior distributions of $\Delta V_0$ and $l$. The radial profiles of the expected local line width, which are computed with the model midplane temperature and input turbulent velocities, are also shown (green curves). In all three cases of different input turbulent velocities without ALMA simulations (Figure \ref{fig:dvprof_RTmodel}a--c), the best-fit local line widths match the expected values mostly within the fitting uncertainties. True and derived turbulent velocities are presented with black and red dashed lines in Figure \ref{fig:dvprof_RTmodel}, respectively. Red shaded regions show the uncertainty of the obtained turbulent velocities, derived by propagating the fitting uncertainty of $\Delta V$. Similarly, in all three cases of different turbulent inputs, the derived turbulent velocities agree with the true input values mostly within uncertainties derived from the statistical fitting errors.

Figure \ref{fig:dvprof_RTmodel}d shows the result for the case of $\vturb=0.5~\cs$ with an ALMA simulation. The deviation of the best-fit local line width (purple curve) from the expected local line width (green curve) slightly increases compared to the case without the ALMA simulation (Figure \ref{fig:dvprof_RTmodel})c. However, the deviation is still a few tens $\mps$ at most. The figure also shows that the turbulent velocity is determined with a similar accuracy. 
Figure \ref{fig:dvprof_RTmodel}e presents the result for the case of $\vturb=0.5~\cs$ without ALMA simulation but with channel masking, in which the same velocity channels excluded in the fitting of the HL Tau data were omitted. The result is nearly identical to obtained without channel masking (Figure \ref{fig:dvprof_RTmodel}c), demonstrating that the local line width can be robustly constrained even when only a subset of velocity channels is used. This is because the spatial extent of the emission in individual channel maps retains information on the line broadening. In addition, in a Keplerian disk, spatial position and LOS velocity are correlated; redshifted and blueshifted emission appears on opposite sides of the disk. As a result, the overall spectral profile can still be recovered at certain spatial locations even when part of redshifted or blueshifted emission is masked.

For further comparison, the radial profile of the local line width derived by fitting the actual \ch{H_2CO} data is also presented with gray curves in Figure \ref{fig:dvprof_RTmodel}. These results show that the large local line width obtained from the data is difficult to reproduce with models with small turbulent velocity of $\vturb \le 0.3\cs$.

As demonstrated, given the same density structure of the disk and S/N comparable with that of the data, the parametric fitting approach provides a reasonable estimate of the nonthermal velocity dispersion with an accuracy of a few tens $\mps$.

\section{Full fitting results for the layered models}\label{app:supp_models}

Table \ref{tab:params_layered} summarizes the full fitting results for the layered models.

% --------- Table ---------
\begin{deluxetable}{llcccc}
%\digitalasset
%\tablewidth{\columnwidth}
\tablecolumns{6}
\tablecaption{Best-fit parameters of the layered models \label{tab:params_layered}}
\tablehead{
\colhead{Parameter} & \colhead{Unit} & \multicolumn{4}{c}{Best-fit value} \\
\colhead{} & \colhead{} & \colhead{$p_z=1$, $z_0$ free} & \colhead{$p_z=2$, $z_0$ free} & \colhead{$p_z=1$, $z/R=0.1$} & \colhead{$p_z=1$, $z/R=0.15$}
}
\startdata
& & \multicolumn{4}{c}{Parameters of the emission height} \\
$z_0$ & au & $4.9 \pm 0.8$ & $3.7 \pm 0.6$ & $10.0$ (fixed) & $15.0$ (fixed) \\
$p_z$ & - & $1.0$ (fixed) & $2.0$ (fixed) & $1.0$ (fixed) & $1.0$ (fixed) \\
$H_0$ & au & $1.5 \pm 0.4$ & $1.8 \pm 0.5$ & $0.8 \pm 0.2$ & $0.6 \pm 0.3$ \\
$p_H$ & - & $0.3 \pm 0.3$ & $0.3 \pm 0.2$ & $0.1 \pm 0.1$ & $0.1 \pm 0.1$ \\
\hline
& & \multicolumn{4}{c}{Other Parameters} \\
$T_\mathrm{g,0}$ & K & $23.8 \pm 0.4$ & $24.0 \pm 0.4$ & $22.9 \pm 0.3$ & $25.3 \pm 0.6$ \\
$q_\mathrm{g}$ & - & $0.59 \pm 0.03$ & $0.58 \pm 0.03$ & $0.57 \pm 0.03$ & $0.62 \pm 0.04$ \\
$\log_{10} N_\mathrm{g,c}$ & $\mathrm{cm}^{-2}$ & $16.0 \pm 0.4$ & $15.9 \pm 0.4$ & $15.7 \pm 0.3$ & $15.9 \pm 0.5$ \\
$R_\mathrm{g,c}$ & au & $91 \pm 8$ & $90 \pm 10$ & $89 \pm 7$ & $90 \pm 8$ \\
$\gamma_\mathrm{g}$ & - & $-0.9 \pm 0.3$ & $-0.8 \pm 0.4$ & $-0.7 \pm 0.3$ & $-0.8 \pm 0.3$ \\
$i$ & $^\circ$ & $45.5 \pm 0.4$ & $45.7 \pm 0.4$ & $46.9 \pm 0.3$ & $48.2 \pm 0.3$ \\
$\theta_\mathrm{PA}$ & $^\circ$ & $135.5 \pm 0.2$ & $135.5 \pm 0.2$ & $135.5 \pm 0.2$ & $135.5 \pm 0.2$ \\
$M_\ast$ & $\Msun$ & $2.33 \pm 0.02$ & $2.32 \pm 0.02$ & $2.27 \pm 0.02$ & $2.25 \pm 0.02$ \\
$\vsys$ & $\kmps$ & $7.238 \pm 0.006$ & $7.238 \pm 0.006$ & $7.246 \pm 0.006$ & $7.258 \pm 0.006$ \\
$\delta x_0$ & au & $0.3 \pm 0.3$ & $0.2 \pm 0.3$ & $0.4 \pm 0.3$ & $0.0 \pm 0.4$ \\
$\delta y_0$ & au & $-3.1 \pm 0.4$ & $-3.1 \pm 0.4$ & $-3.4 \pm 0.4$ & $-2.9 \pm 0.4$ \\
$\Delta V_0$ & $\kmps$ & $0.16 \pm 0.01$ & $0.16 \pm 0.01$ & $0.15 \pm 0.01$ & $0.096 \pm 0.007$ \\
$l$ & - & $-0.4 \pm 0.2$ & $-0.3 \pm 0.1$ & $-0.3 \pm 0.1$ & $-0.6 \pm 0.2$ \\
\enddata
\tablecomments{Parameters of the dust layer are set to the same as those for the fiducial model, which are summarized in Table \ref{tab:params}.}
\end{deluxetable}

\section{Supplementary Material for Model Fits}\label{app:supp_fig}

% ------------ Figure -------------
\begin{figure*}
   \centering    \includegraphics[width=0.9\linewidth]{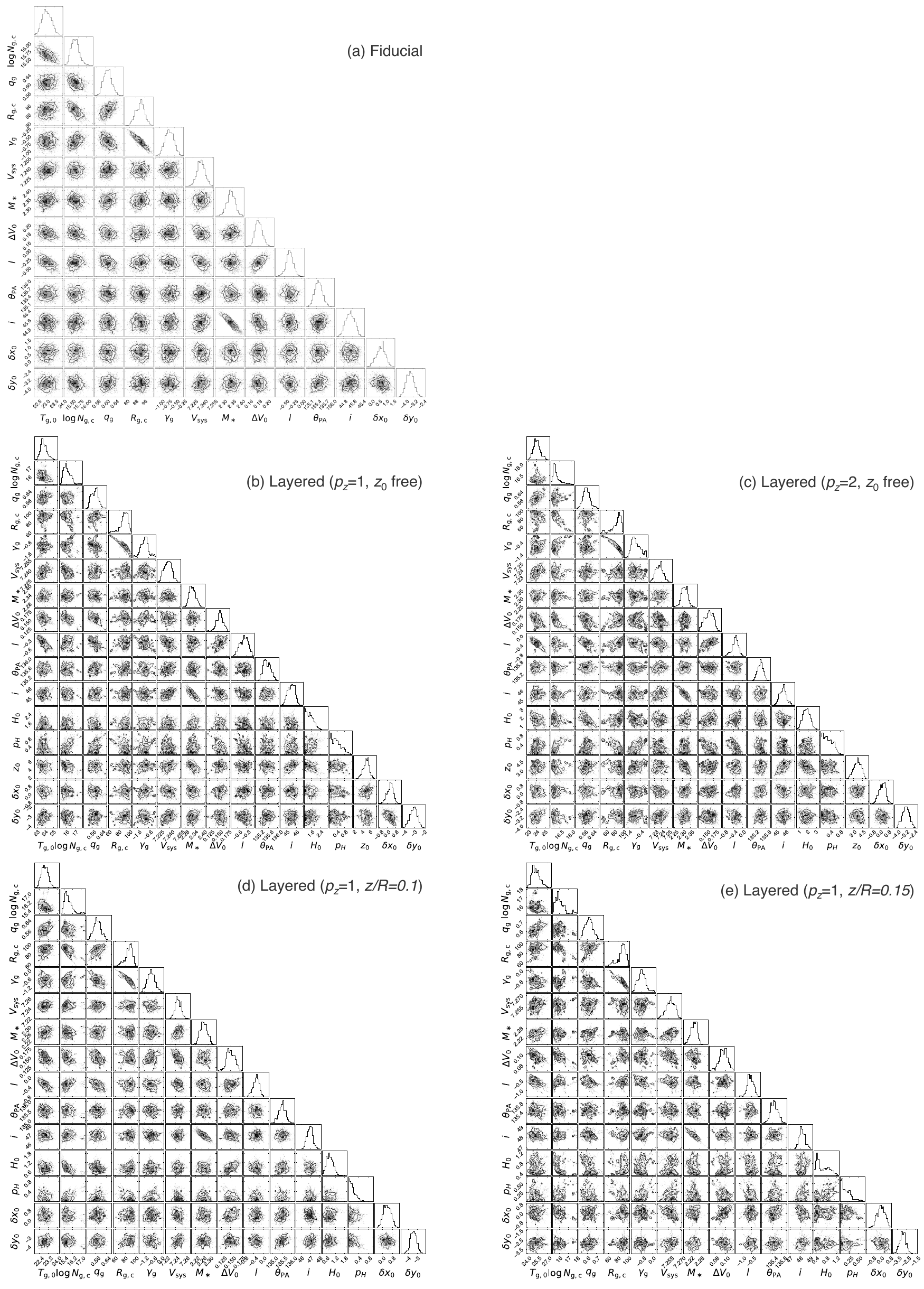}
    \caption{Corner plots showing the posterior distributions for all model fits.}
    \label{fig:corner}
\end{figure*}
% ----------------------------------

The corner plots of the posterior distributions for all model fits are presented in Figure \ref{fig:corner}.

%%% Bibliography
\bibliography{ref_HLTau_paper}{}
\bibliographystyle{aasjournalv7}

\end{document}